\documentclass[10pt]{iopart}
\pdfoutput=1
 		            		
\usepackage{iopams}  		       
\usepackage{graphicx}									
\usepackage{caption}
\usepackage{subcaption}
\usepackage{url}
\usepackage{hyperref}
\usepackage{color,soul}
\usepackage{cite}

\newcommand{\bq}{\boldsymbol{q}}
\newcommand{\bx}{\boldsymbol{x}} 
\newcommand{\expval}[1]{\left \langle #1 \right \rangle}
\newcommand{\curl}{\bnabla \times}
\renewcommand{\div}{ \bnabla \boldsymbol{\cdot} }

\begin{document}

\title{Steady state entropy production rate for scalar Langevin field theories}
\author{Yuting I. Li, Michael E. Cates}

\address{DAMTP, Centre for Mathematical Sciences, Wilberforce Rd, Cambridge CB3 0WA}
\ead{yuting.li@damtp.cam.ac.uk}

\begin{abstract}

The entropy production rate (EPR) offers a quantitative measure of time reversal symmetry breaking in non-equilibrium systems. It can be defined either at particle level or at the level of coarse-grained fields such as density; the EPR for the latter quantifies the extent to which these coarse-grained fields behave irreversibly. In this work, we first develop a general method to compute the EPR of scalar Langevin field theories with additive noise. This large class of theories includes active versions of Model A (non-conserved density dynamics) and Model B (conserved) and also models where both types of dynamics are simultaneously present (such as Model AB \cite{modelAB}). Treating the scalar field $\phi$ (and its time derivative $\dot\phi$) as the sole observable(s), we arrive at an expression for the EPR that is non-negative for every field configuration and is quadratic in the time-antisymmetric component of the dynamics. Our general expression is a function of the quasipotential, which determines the full probability distribution for configurations, and is not generally calculable. To alleviate this difficulty, we present a small-noise expansion of the EPR, which only requires knowledge of the deterministic (mean-field) solution for the scalar field in steady state, which generally is calculable, at least numerically. We demonstrate this calculation for the case of Model AB \cite{modelAB}. We then present a similar EPR calculation for Model AB with the conservative and non-conservative contributions to $\dot\phi = \dot\phi_{\rm A} + \dot\phi_{\rm B}$ viewed as separately observable quantities. The results are qualitatively different, confirming that the field-level EPR depends on the choice of coarse-grained information retained within the dynamical description.

\end{abstract} 

\maketitle 

\section{Introduction} 
Non-equilibrium physics is ubiquitous. Examples range from biological systems where energy is consumed and dissipated on the microscopic scale \cite{MIPS,  ramaswamy2010, marchetti2013review}, to driven diffusive systems where some external macroscopic force drives the system away from equilibrium \cite{schmittmann1995, evans1998}. Recently there has been some interest in quantifying the deviation from equilibrium, and entropy production is such a quantity \cite{FodorPRL, nardini2017entropy, seifert, pietzonka2017entropy, seifert2005entropy, MandalEntropyActive, ShankarMarchetti, niggemann2020}. 

In a seminal paper by Seifert \cite{seifert2005entropy}, he introduced the notion of entropy production for a single trajectory, defined in terms of the probability of the forward and the backward path. The total entropy production can be split into two parts: the entropy production of the system and that of the medium. The entropy production of the system gives the Gibbs entropy $\sum_\mathrm{states} p_n \log p_n$ upon averaging over the probability distribution, whereas the medium entropy production is directly linked to the heat production in simple cases. Since then, there have been various work on particle-based entropy production in the context of active matter \cite{DabelowPRX, MandalEntropyActive, ShankarMarchetti, FodorPRL}.  

From an informatic point of view, the entropy production is a quantitative answer to the question: given certain information about the system, how irreversible does the dynamics appear? Sometimes the strong non-equilibrium nature of the underlying microscopic interactions (e.g.\ self propulsion) do not survive the coarse graining procedure \cite{pietzonka2017entropy} -- the dynamics can appear equilibrium-like on a scale much larger than the individual agents \cite{MIPS,  TjhungPRX2018, modelAB}. To address this, Nardini et al.\ extended the definition of the entropy production to field theories, quantifying the amount of time reversal symmetry breaking at the macroscopic level \cite{nardini2017entropy}. They proposed an expression for the rate of entropy production of Active Model B (a minimal way to add non-equilibrium perturbations to Hohenberg and Halperin's Model B \cite{HH, MIPS}) and computed its steady-state values via numerical average of stochastic trajectories. 

Meanwhile, in the Macroscopic Fluctuation Theory literature, the concept of symmetric and antisymmetric currents were introduced for diffusive systems, as another way to pinpoint the irreversible aspect of the dynamics \cite{mft}. Bertini el al.\ constructed an adjoint system whose forward time evolution is identical to a movie of the original system played backwards. Then the antisymmetric current corresponds to the difference between the original current and the current of the adjoint system.

In this work, we investigate the link between the entropy production rate and the antisymmetric current. We build upon the results of Nardini et al.\ \cite{nardini2017entropy} and study the instantaneous field-theoretic entropy production rate (EPR) of the general class of scalar field theories with additive noise. This includes equilibrium Model A and Model B, which were systematically catalogued in Hohenberg and Halperin \cite{HH}, as well as their non-equilibrium extensions, such as driven diffusive models \cite{mft} and Active Model B+ \cite{MIPS, TjhungPRX2018}. Finally we use Model AB \cite{modelAB} as a case study on the effect of tracking different information. Model AB describes systems with separate conservative and non-conservative component, often driven by different underlying mechanisms. In our previous work \cite{modelAB}, we found a special subspace where time reversal symmetry is apparently restored for the density field, prompting questions on the behaviour of different EPRs in and near this equilibrium subspace. 

The paper is organised as follows. In section \ref{entropy_production}, we introduce the entropy production associated with a trajectory, as well as its decomposition in terms of internal entropy production and external entropy production, analogous to the system and medium splitting of Seifert \cite{seifert2005entropy}. Section \ref{scalar_langevin} catalogues various scalar Langevin systems and their steady state solutions, including a detailed discussion Model AB and its special equilibrium subspace. Next, in section \ref{phi_dot_epr}, we compute the entropy production rate and its connection to the antisymmetric component of the dynamics (defined in a similar way to the antisymmetric current in Macroscopic Fluctuation Theory \cite{mft}). A small noise expansion of the EPR is also presented to make progress on models that are not exactly solvable and we demonstrate the method with an application to Model AB. In section \ref{phi_dot_a_epr}, we investigate the effect of tracking different information by keeping account of separate Model A and Model B contributions in Model AB, and compare the resulting EPR with the results of tracking $\phi$-evolution only. In the final section, the results are summarised and possible future work is proposed.

\section{Entropy production} 
\label{entropy_production}
By the macroscopic nature of field theories, a trajectory or path in the space of field configurations is a bundle of all microscopic realisations that give rise to the same coarse grained description. The segregation into bundles depends on what macroscopic variables (such as the density of a species, the composition variable of a binary fluid or the local particle density) are tracked -- broadly speaking, more information leads to finer bundles. The entropy production $\Delta S$ associated with such a trajectory is, according to stochastic thermodynamics, defined as the log of the ratio of the probability of the forward trajectory and the backward trajectory  \cite{seifert2005entropy, nardini2017entropy}, 
\begin{equation}
\Delta S = \log \frac{ \mathbb{P} [\mathrm{all \ microscopic \ realisations \ of \ the \ forward \ path}] }{\mathbb{P}[\mathrm{all \ microscopic \ realisations \ of \ the \ backward \ path}] }
\end{equation}
Thus it is impossible to talk about entropy production without specifying the information being tracked.  Denote the set of available macroscopic variables as $ \left \{\boldsymbol{X} (\bx, t)_{t \in [0, \tau]} \right \}$, where the components of $\boldsymbol{X}$ are the variables tracked, $(\bx, t)$ denotes the dependence of $\boldsymbol{X}$ on space and time, and $\tau$ is the length of the trajectory. Following the stochastic thermodynamics literature, we define an intermediate quantity, the rate function  $\mathbb{R}$, as the $\log$ of the path probability $\mathbb{P}$ \cite{mft, seifert}, 
\begin{equation}
\mathbb{P} \left [ \{\boldsymbol{X}(\bx, t)_{t \in [0, \tau]} \} \right ] \propto \exp \left ( -\epsilon^{-1}  \mathbb{R} \left [\{\boldsymbol{X}(\bx, t)_{t \in [0, \tau]} \}  \right ] \right )
\label{eq:rate_function}
\end{equation}
where $\epsilon$ is a parameter that quantifies the amount of the noise (the meaning will become clear when we write down the rate function explicitly for a scalar Langevin system). Notation-wise, throughout this paper we use $\mathbb{F}$ to denote functionals of a space-time trajectory, $\mathcal{F}$ for a functional of a spatial field configuration and $F$ to denote functions. We will also silently omit the $\bx$-dependence and the explicit time dependence of the trajectories from now on. For example, $\mathbb{P}\left [\{ \boldsymbol{X} \} \right ]$ is the probability of the a specific realisation of the time evolution of the fields from time $0$ to $\tau$, whereas $\mathcal{P}[\boldsymbol{X}(t), t]$ is the probability of observing the configuration $\boldsymbol{X}(t)$ at time $t$. 

Next, we further factorise the path probability $\mathbb{P}\left [\{ \boldsymbol{X} \} \right ]$ into a product of the probability of the initial conditions $\mathcal{P}[ \boldsymbol{X}(0), 0]$ and the conditional probability of the evolution given the initial conditions $\mathbb{P} \left [  \{\boldsymbol{X} \} \middle \vert  \boldsymbol{X}(0) \right ]$, 
\begin{equation}
\mathbb{P}\left [ \{\boldsymbol{X} \} \right ] = \mathcal{P} \left [ \boldsymbol{X}(0), 0 \right ] \mathbb{P} \left [  \{\boldsymbol{X} \} \middle \vert  \boldsymbol{X}(0)\right ] 
\end{equation}
Taking the logarithms of the both sides yields the splitting of the rate function $\mathbb{R}$ in terms of an ``instantaneous quasipotential $\mathcal{V}$'' (this is not the conventional definition of quasipotential \cite{mft} though they do coincide in steady state, which we denote as $\mathcal{V}_\mathrm{ss}$) and the action $\mathbb{A}$ of the trajectory,  
\begin{equation}
\eqalign{  
\mathbb{R}\left [ \{\boldsymbol{X}\} \right ] = \mathcal{V}\left [\boldsymbol{X}(0), 0 \right ] + \mathbb{A} \left [  \{\boldsymbol{X} \}\right ] \\
\mathcal{P}[\boldsymbol{X}(t), t] \propto \exp \left ( - \epsilon^{-1} \mathcal{V}[  \boldsymbol{X}(t), t] \right ) \\
\mathbb{P}\left [\{\boldsymbol{X}\} \middle \vert  \boldsymbol{X}(0) \right ] \propto \exp \left  ( - \epsilon^{-1} \mathbb{A}[ \{\boldsymbol{X} \}  ] \right )
} 
\label{eq:a_v}
\end{equation}
where we note that the explicit $t$-dependence in $\mathcal{V}, \mathcal{P}$ highlights the fact that the probability distribution $\mathcal{P}$ can change over time. In equilibrium, $\epsilon^{-1} \mathcal{V}_\mathrm{ss} = \beta \mathcal{F} $ where $\mathcal{F}$ is the (mesoscopic) free energy functional and $\beta = 1/k_\mathrm{B} T$ is the inverse temperature. 

In stochastic thermodynamics, the entropy production of a trajectory $\Delta S \left [ \{ \boldsymbol{X} \} \right ]$ is defined as proportional to the difference between the rate function for the forward trajectory $\left \{\boldsymbol{X}(t)_{t \in [0, \tau]}  \right \} $ and that of the backward trajectory $\left \{\boldsymbol{X}^\mathrm{R} (t)_{ t \in [0, \tau]} \right \}$ \cite{seifert2005entropy, seifert}. The backward time evolution is related to the forward one in a rather intuitive way: $X_\alpha^\mathrm{R}(t) = \theta_\alpha X_\alpha(\tau - t)$ where $\alpha$ denotes the index and $\theta_\alpha = \pm 1$ depending on whether the variable is even or odd under time reversal (e.g.\ current is odd, density is even) \cite{seifert, mft}, 
\begin{equation}
\epsilon \Delta S[ \{\boldsymbol{X}\} ] = - \mathbb{R}[ \{\boldsymbol{X}\} ] + \mathbb{R}[\{\boldsymbol{X}^\mathrm{R} \} ] 
\end{equation}
There is a technical detail associated with the path reversal: the action must be written with Stratonovich (midpoint) discretisation \cite{seifert, lau} so that the time reversal of the trajectory has the same time discretisation as the forward trajectory \cite{nardini2017entropy}. Throughout this paper, we adopt the Stratonovich discretisation scheme (except one occasion in appendix \ref{ap:exp} where another choice is explicitly stated). 

Similarly to the splitting of the rate function, the entropy production can be decomposed into an internal $ \Delta S_\mathrm{Int}[\{\boldsymbol{X} \} ] $, that only depends on the quasipotentials $\mathcal{V}$, and an external part $\Delta S_\mathrm{Ext}[ \{\boldsymbol{X} \} ]$, which depends on the actions $\mathbb{A}$, 
\begin{equation}
\eqalign{  
\Delta S[\{\boldsymbol{X}\} ]  &= \Delta S_\mathrm{Int}+ \Delta S_\mathrm{Ext} \\
\epsilon \Delta S_\mathrm{Int}[\{\boldsymbol{X} \} ] &= \mathcal{V}[ \boldsymbol{X}(\tau), \tau ] - \mathcal{V} [  \boldsymbol{X}(0) , 0] \\ 
\epsilon \Delta S_\mathrm{Ext}[ \{\boldsymbol{X} \} ] &= - \mathbb{A}[ \{ \boldsymbol{X} \} ] + \mathbb{A}[ \{ \boldsymbol{X}^\mathrm{R} \} ] 
} 
\label{eq:int_ext}
\end{equation}
The internal entropy production is the field-theoretic extension of Seifert's entropy production of the system, both of which are the difference between the initial and final quasipotential $\mathcal{V}$ \cite{seifert2005entropy}. Seifert's argument for the connection to Gibbs entropy also carries through: taking the ensemble average of $\Delta S_\mathrm{Int}$, 
\begin{equation} 
\eqalign{  
\fl \expval{ \Delta S_\mathrm{Int}} = \frac{1}{\epsilon} \expval{\mathcal{V}(\tau)}_{\mathcal{P}(\tau)} - \frac{1}{\epsilon} \expval{ \mathcal{V}(0)}_{\mathcal{P}(0)} \\
\fl \qquad \quad= - \int \prod_\alpha \mathcal{D}X_\alpha \left [ \mathcal{P}( \tau)\log \mathcal{P}(\tau) \right ]  + \int \prod_\alpha \mathcal{D}X_\alpha \left [ \mathcal{P}(0)\log \mathcal{P}( 0) \right ] + C 
} 
\end{equation} 
where $C$ is a constant and $\int \prod_\alpha \mathcal{D}X_\alpha$ represents the integration over all configurations of $ \boldsymbol{X} $. Observe that the last two terms are the Gibbs entropy of the final and the initial configurations respectively, implying that $\expval{ \Delta S_\mathrm{Int}}$ can be interpreted as the change in the (system) Gibbs entropy. 

On the other hand, the meaning of the external entropy production is less clear, though the choice is unique once the internal part of the entropy production is identified. For the example of an over-damped particle introduced in Seifert's paper \cite{seifert2005entropy}, the external entropy production can be directly related to the heat dissipated in the bath. However, the coarse graining of the particle dynamics into field trajectories changes the amount of information tracked, and hence the observed entropy production \cite{pietzonka2017entropy}. As a consequence, any direct link between the entropy production and physical, as opposed to informatic, quantities can only be established in a model-specific way \cite{DabelowPRX}, if such interpretations exist at all. We would like to note that we are addressing the informatic view of entropy production in this work, rather than heat flow, and we refer to Markovich et al.\ for treatments of the latter in field theories \cite{Markovich2020}. 

Going back to the calculation, the instantaneous internal (resp.\ external) entropy production rate (EPR) can be obtained by differentiating the internal (resp.\ external) entropy production with respect to the final time $\tau$. The expression for the general case is rather cumbersome and not particularly enlightening, so we will only present the formula for the choices of  $\boldsymbol{X}$ of our interest. Note that our approach is consistent with the method in Nardini et al.\ \cite{nardini2017entropy}: $S$ in equation (16) of their paper corresponds to our $\Delta S_\mathrm{Ext}$, but as they subsequently divide by $\tau$ and take $\tau \rightarrow \infty$ in their definition of the EPR, the internal entropy production $\Delta S_\mathrm{Int}$, an $O(\tau^0)$ piece, vanishes in the process. 

Having developed a general scheme for calculating the entropy production rate of stochastic trajectories, we proceed to introduce the specific class of systems that the formulation will be applied to in the remainder of this paper. 

\section{Scalar Langevin systems} 
\label{scalar_langevin} 
In this paper we focus our attention on scalar Langevin systems with additive white noise. Extensions to vectorial systems or systems with multiplicative noise are possible but bring additional complications \cite{lau}, which we will not discuss here. In this section, we will first write down the most general form, and follow with some examples, including relaxational models (as defined in Tauber et al. \cite{tauber2014})  and their non-equilibrium extensions. We will also introduce non-equilibrium Model AB, a model we proposed in our previous paper \cite{modelAB} for phase separating systems with additional non-equilibrium reactions. 

Consider a scalar field $\phi$, which can be the (rescaled) density of some particles or the composition variable of a binary fluid. The most general form of Langevin dynamics is, 
\begin{equation}
\eqalign{  
&\partial_t \phi  = F (\phi) + \sqrt{2 \epsilon} \sigma \Lambda\\
&\expval{\Lambda( \bx, t) \Lambda(\boldsymbol{y}, s)} = \delta (\bx - \boldsymbol{y}) \delta (t -s) 
} 
\label{eq:langevin}
\end{equation}
where $F(\phi)$ is the deterministic dynamics (not to be confused with free energy, which we denote as $\mathcal{F}[\phi]$ as will be specified later), $\Lambda$ is a spatial-temporal white noise, $\sigma$ is an operator independent of $\phi$ (to be defined later) and $\epsilon$ characterises the noise strength. We note that $\epsilon$ is the same as the previously mentioned constant in the definition of the rate function, the action and the quasipotential in equation (\ref{eq:rate_function}, \ref{eq:a_v}). In equilibrium thermodynamics, $\epsilon = k_\mathrm{B} T$, whose role is usually singled out by convention and not absorbed in the definition of quantities such as free energy.  

Before we define $\sigma$, we take a detour to introduce the notations we use for fields and operators in this paper. Adopting the notation from linear algebra, we treat the scalar field $\phi$ as an infinite dimensional column vector and define its adjoint $\phi^\dagger$ as the corresponding row vector. This enables us to proceed in a basis-independent way and only refer to a specific basis (such as real space or Fourier space) when needed. The inner product between scalar fields $\phi$ and $\psi$, denoted as $\phi^\dagger \psi$, is the sum of the products of their elements with respect to some an orthogonal basis: $\phi^\dagger \psi = \sum_{i=1}^\infty \phi_i^* \psi_i$. For example, in real space $\phi^\dagger \psi = \int \mathrm{d} \bx\, \phi (\bx) \psi (\bx)$, assuming $\psi(\bx), \phi(\bx)$ are real;  in Fourier space the inner product is $(2 \pi)^{-d} \int \mathrm{d} \bq \, \phi(- \bq) \psi(\bq)$ (note we take the Fourier transform convention $\phi(\boldsymbol{x} ) = (2 \pi)^{-d} \int \mathrm{d} \bq \phi(\bq ) \exp( - i \bx \boldsymbol{\cdot} \bq ) $ in d dimensions). We can similarly define the outer product in an element-wise way, $(\phi \psi^\dagger)_{ij} = \phi_i \psi_j^*$, naturally extending from vectors to matrices. The operation of matrix $O$ on vector $\phi$ is defined in the usual way as $(O \phi)_i = \sum_{j=0}^\infty O_{ij} \phi_j$. This leads to the definition of the adjoint of a matrix, denoted as $O^\dagger$, defined element-wise as $O^\dagger_{ij} = O_{ji}^*$, with the property that $\phi^\dagger (O \psi) = (O^\dagger \phi)^\dagger \psi$. 

With these notations, $\sigma$ is, in general, an infinitely dimensional matrix that is not a function of $\phi$, as we assumed that the noise is not multiplicative. The spatial-temporal noise $\Lambda$ is a vector and its correlation can be denoted by an outer product: $\expval{\Lambda(t) \Lambda(s)^\dagger} = I \delta(t - s)$ where $I$ is the identity, e.g.\ in real space $I$ is the delta function $ \delta(\bx - \boldsymbol{y})$. Letting $\eta(t) = \sigma \Lambda(t)$, the correlation of $\eta$ is 
\begin{equation}
\expval{ \eta(t) \eta^\dagger(s) } = \expval{ (\sigma \Lambda(t))(\sigma \Lambda(s))^\dagger} = \sigma \sigma^\dagger \delta( t- s) \equiv K \delta( t- s)
\end{equation}  
where we have defined $K = \sigma \sigma^\dagger$, commonly known as the noise kernel. Crucially, $\sigma$ is only defined so far as $\sigma \sigma^\dagger$ yields the desired noise kernel $K$, because noises with the same mean and correlation are indistinguishable \cite{gardiner}. For convenience we will always choose a specific form of $\sigma$ with the understanding that many other choices are equivalent, as will be illustrated in the section below. 

\subsection{Relaxational models and non-equilibrium modifications} 
\label{relax}

A well-studied subclass of scalar Langevin dynamics consists of the relaxational models systematically catalogued in Hohenberg and Halperin's review \cite{HH} that describes the dynamical approach to equilibrium, adopting a top-down method that classifies models based on the symmetries and conservation laws. For a single scalar field, the dynamics is named Model B if the $\phi$ field is conserved locally (i.e. $\partial_t \phi = \div \boldsymbol{J}$ for some current $\boldsymbol{J}$) or Model A for the non-conservative case.  With the formalism introduced in the previous section, Model A and Model B can be concisely written down as follows (the noise kernels are diagonal in Fourier space so only the diagonal elements are presented), 
\begin{equation}
K_\mathrm{X}(\bq) =M_\mathrm{X} |\bq|^{2 \lambda_\mathrm{X}},  \qquad F_\mathrm{X} = - K_\mathrm{X} \frac{\delta \mathcal{F}_\mathrm{X}}{\delta \phi}, \quad
\qquad \mathrm{for \ X} = \mathrm{A, B}
\label{eq:modelA,modelB}
\end{equation}
where $M_\mathrm{X}$ is a mobility constant and $\lambda_\mathrm{A} = 0, \lambda_\mathrm{B} = 1$. Alternatively, we can write the noise kernel in real space: $K_\mathrm{X} = M_\mathrm{X} ( i \nabla)^{2 \lambda_\mathrm{X}}$, which is no longer diagonal (see \ref{ap:num} for the explicit form of the discrete Laplacian operator on a lattice). As discussed before, the $\sigma$ matrix is ambiguous, and here we choose it to also be diagonal in Fourier space with elements $\sigma_\mathrm{X}(\bq) = M_\mathrm{X} ( - i | \bq|)^{\lambda_\mathrm{X}}$. The factor of $ - i$ in the definition of $\sigma$ is picked such that in 1D, $\sigma_\mathrm{B} = \sqrt{M_\mathrm{B}} \partial_x$. We do note that the more popular representation of noise in Model B is $\div \boldsymbol{\Lambda}$ where $\boldsymbol{\Lambda}$ is a vectorial white noise. This is equivalent to our definition here because $(- i \bq \cdot \boldsymbol{\Lambda}(\bq))$ has the same noise correlation as $(- i | \bq | \Lambda(\bq))$. 

For both Model A and Model B, it can be shown that the Fokker-Planck equation for the probability $\mathcal{P}[\phi, t]$ evolves towards the Boltzmann distribution with free energy $\mathcal{F}_\mathrm{X}[\phi]$ and temperature $k_\mathrm{B}T = \epsilon$. These free energies are generally chosen to be of $\phi^4$ square-gradient form -- see next section. Furthermore, the system has time reversal symmetry and the $(\phi, \partial_t \phi)$ trajectories obey the principle of detailed balance: the probability of observing a trajectory is the same as the probability of observing the same trajectory in reverse \cite{mukamel, ChaikinLubensky, MikeLesHouchesNotes}. In fact, the same conclusions hold for any sufficiently well-behaved $K$, as long as there is a ``chemical potential'' $\mu$ such that (i) $F = - K \mu$ and (ii) $\mu$ can be written as a functional derivative of some free energy $\mathcal{F}$. 

These relaxational models can be extended to non-equilibrium in a `minimal' way by adding terms that cannot be absorbed via a modification of the free energy (i.e. $ \mu \neq \delta_ \phi \mathcal{F} $). In particular, in systems with conservation law, we can add a driving term $\boldsymbol{E}(\phi, x)$ to Model B, \begin{equation}
\partial_t \phi = M_\mathrm{B}  \div  \left ( \bnabla \frac{\delta \mathcal{F}_\mathrm{B} }{\delta \phi} - \boldsymbol{E} \right ) + \sqrt{2 \epsilon} \sigma_\mathrm{B} \Lambda  
\label{eq:driven_diffusive}
\end{equation}
As long as $\boldsymbol{E}$ cannot be written as  $(- \bnabla \mu)$ for some truly integrable chemical potential $\mu$, the system no longer obeys detailed balance and there is no time reversal symmetry \cite{MIPS, mft}. There is generically no analytical solution for the steady state distribution, though in a few special settings it is possible to map back to an equilibrium system. One such example is a 1D system with periodic boundary conditions and $E = \gamma \phi$, modelling interacting particles driven around a ring by a constant force parallel to the ring\footnote{The full stochastic PDE for such a system would be $\partial_t \rho = \partial_x  \left [ D \rho \left ( \partial_x \mu + \gamma \right ) + \sqrt{ 2 D \rho } \Lambda \right ]$. Perturb around some constant density $\rho = \rho_0 ( 1 + \phi)$ and we get the one written here once we omit the $\phi$-dependence in the mobility. }. After a Galilean transformation $\phi(x, t) \rightarrow \phi(x -M_\mathrm{B} \gamma t, t)$, the equation is identical to Model B, hence the steady state quasipotential $\mathcal{V}_\mathrm{ss}$ is the same as the free energy $\mathcal{F}_\mathrm{B}$ in this case.

A popular choice in active matter literature is to make $\boldsymbol{E}$ depend on $\phi$ in a way that breaks the time reversal symmetry to lowest order in $\phi$ and $\bnabla$ \cite{MIPS, toner_tu, TjhungPRX2018}. For mass-conserving systems, the lowest order terms are $\boldsymbol{E}(\phi) = \lambda \bnabla | \bnabla \phi |^2 + \zeta (\bnabla \phi) (\nabla^2 \phi)$, named Active Model B+ in Tjhung et al.\  \cite{TjhungPRX2018} (the third term of $O(\nabla^3 \phi^2)$ can be absorbed into the free energy if $\lambda$ and $\zeta$ are also adjusted). The former, the $\lambda$-term, is a gradient of a local chemical potential $\lambda | \bnabla \phi |^2$, which cannot be written as a functional derivative of any free energy. The latter, the $\zeta$-term, leads to macroscopic steady state current, though only its curl-free piece contributes to the $\phi$-dynamics as argued in Tjhung et al.\ \cite{TjhungPRX2018}. We will expand upon this argument when we look at the entropy production of non-equilibrium systems with conservation laws.

\subsection{Non-equilibrium Model AB} 
\label{modelAB}
Another novel way of breaking time reversal symmetry, that has generated a lot of interest recently, is to combine two equilibrium dynamics with different free energies \cite{active_emulsions, modelAB, grafkePRL2017, pattern, Glotzer1, Glotzer2, GlotzerPRE, zwickerNatPhys2017, ZwickerPRE, puri1998phase}. Model AB, constructed from Model B and Model A as the name suggests, represents systems with one scalar field $\phi$, subject to \textit{separate} diffusive and reactive dynamics, such as binary fluids with non-equilibrium chemical reactions and active particles with population dynamics \cite{modelAB}. Recall the definitions of Model A and Model B in equation (\ref{eq:modelA,modelB}) and add contributions from both, 
\begin{equation}
\eqalign{  
\partial_t \phi &= \partial_t \phi_\mathrm{A} + \partial_t \phi_\mathrm{B} \\
\partial_t \phi_\mathrm{A} &=  - M_\mathrm{A} \mu_\mathrm{A} (\phi) + \sqrt{2 \epsilon M_\mathrm{A}} \Lambda_\mathrm{A} \\
\partial_t \phi_\mathrm{B} &= - K_\mathrm{B} \mu_\mathrm{B} (\phi) + \sqrt{2 \epsilon} \sigma_\mathrm{B} \Lambda_\mathrm{B} \\ 
} 
\label{eq:phi_dot_ab}
\end{equation}
where $\Lambda_\mathrm{A}, \Lambda_\mathrm{B}$ are independent unit white noises and recall that $\sigma_\mathrm{B}(\boldsymbol{q}) = - i |\bq | \sqrt{M_\mathrm{B}}, K_\mathrm{B}(\bq) = M_\mathrm{B} | \bq |^2 $. The two chemical potentials $\mu_\mathrm{A,B}$ are not necessarily functional derivatives of free energies $\mathcal{F}_\mathrm{A, B}$. For simplicity, we assume here, as in our previous paper \cite{modelAB}, that they are, but allow $\mathcal{F}_\mathrm{A} \neq \mathcal{F}_\mathrm{B}$. This is enough to break time-reversal symmetry, and generically does so at lower order in $(\nabla, \phi)$ than the terms needed to break the symmetry in either sector by itself. 

Although this set of equations appears different from equation (\ref{eq:langevin}) at first sight, they can be rewritten in that form by combining the two Gaussian noises into one. Let $\eta = \sqrt{M_\mathrm{A}} \Lambda_\mathrm{A} + \sigma_\mathrm{B} \Lambda_\mathrm{B}$ and we want to find a suitable $\sigma$ such that $\eta = \sigma \Lambda$ for a unit white noise $\Lambda$. As argued in section \ref{scalar_langevin}, $\sigma$ can be found by effectively `square rooting' the noise kernel $K$, which is also the spatial factor\footnote{the spatial-temporal correlation factorises into spatial and temporal contributions as shown in the equation below.} of the noise correlation $\expval{\eta(t) \eta^\dagger(s)}$, 
\begin{equation}
\eqalign{
 \expval{ \eta(t) \eta(s)^\dagger} &= M_\mathrm{A} \expval{\Lambda_\mathrm{A}(t) \Lambda_\mathrm{A}(s)^\dagger} + M_\mathrm{B} \expval{ \left ( \sigma \Lambda_\mathrm{B}(t) \right ) \left (\sigma \Lambda_\mathrm{B}(s) \right )^\dagger }  \\
&= \left ( M_\mathrm{A} I - M_\mathrm{B} \nabla^2 \right ) \delta(t - s) 
}
\end{equation}
where we have used the fact that the two noises $\Lambda_\mathrm{A, B}$ are independent. We can now read off the noise kernel $K =   M_\mathrm{A} I - M_\mathrm{B} \nabla^2$, which is diagonal in Fourier space with elements $K(\boldsymbol{q}) = M_\mathrm{A} + M_\mathrm{B} | \bq |^2$. Choosing $\sigma$ to also be diagonal in Fourier space for convenience, then $\sigma(\boldsymbol{q})$ can be any complex root of $K(\boldsymbol{q})$. Here we set $\sigma(\boldsymbol{q}) =  \sqrt{M_\mathrm{A}} - i | \bq | \sqrt{M_\mathrm{B}}$, such that in 1D, $\sigma(x)= \sqrt{M_\mathrm{A}} + \sqrt{M_\mathrm{B}} \partial_x $. Collecting the terms, in the form of equation (\ref{eq:langevin}), Model AB can be written as,  
\begin{equation}
\partial_t \phi = M_\mathrm{B} \nabla^2 \mu_\mathrm{B} - M_\mathrm{A} \mu_\mathrm{A} + \sqrt{2 \epsilon} \sigma \Lambda 
\label{eq:modelab} 
\end{equation}

We showed in our previous paper \cite{modelAB} that there is a special subspace where the $\phi$-dynamics is effectively equilibrium. Since $K$ is non-singular (it is diagonal in Fourier space with nonzero eigenvalues), the inverse exists and we can always find $\mu = - K^{-1} F = \mu_\mathrm{B} + M_\mathrm{A} K^{-1}(\mu_\mathrm{A} - \mu_\mathrm{B})$.  As discussed in the section \ref{relax}, if there exists a $\mathcal{F}$ that has this $\mu$ as its functional derivative, time reversal symmetry will be restored for the $\phi$ field. This includes the trivial ``true equilibrium'' case: $\mu_\mathrm{A} = \mu_\mathrm{B}$, where the phase separation and the chemical reactions are governed by same underlying equilibrium chemical potentials \cite{active_emulsions}. 

\begin{figure}
\centering
\includegraphics[width=0.55\textwidth]{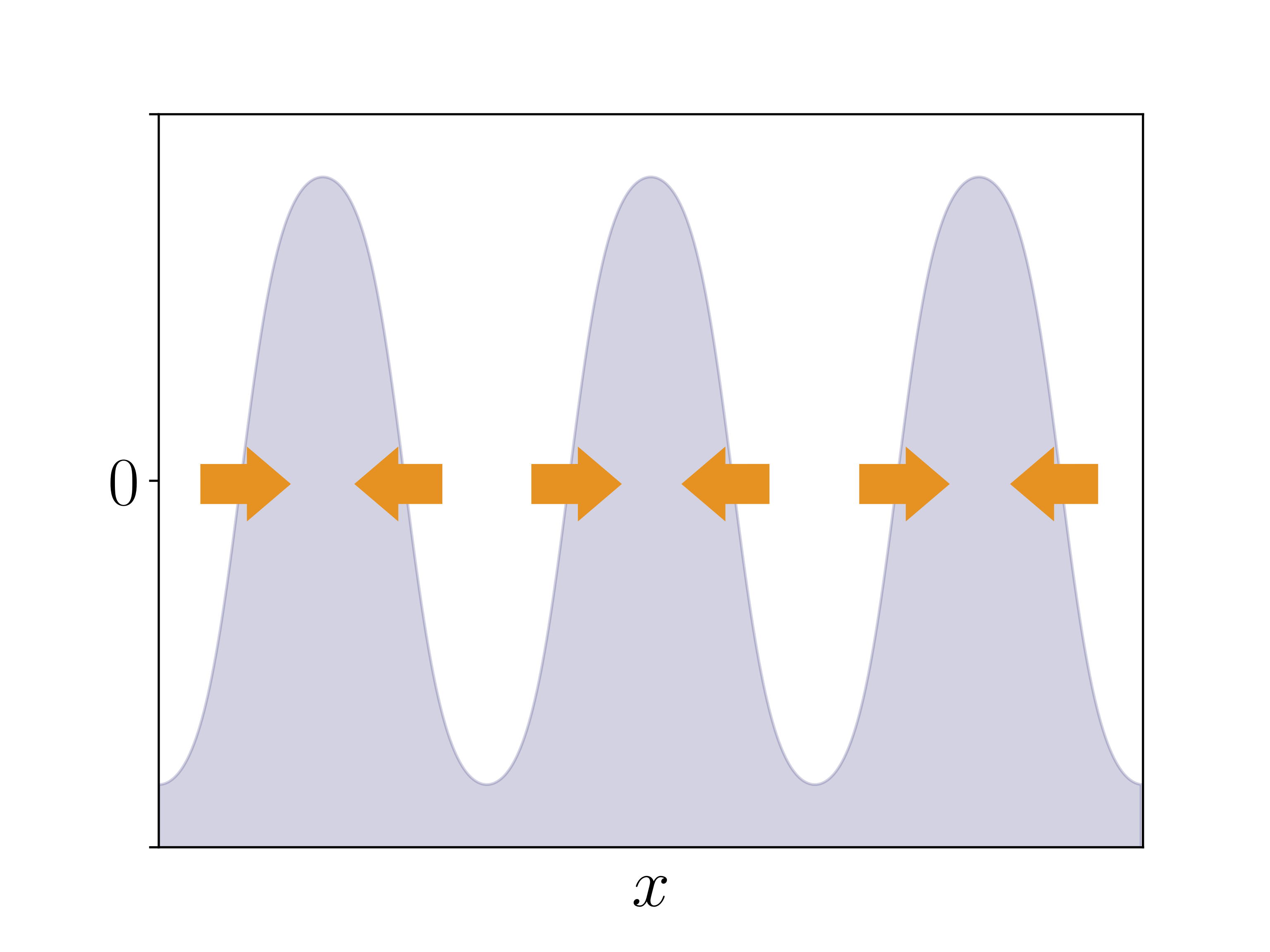}
\caption{Arrested phase separation in one dimension. The orange arrows indicate the steady state currents from dilute regions, where particles are created, to dense regions, where particles are annihilated. In the equilibrium subspace ($\beta' = \beta$), a movie of $\phi(\bx, t)$ would show time symmetric dynamics for the density field, whereas the currents remain large and irreversible. }
\label{fig:steadystate}
\end{figure}

A more general sufficient condition is when $\mu_\mathrm{B} - \mu_\mathrm{A} = K Q \phi$ for any self-adjoint matrix $Q$ independent of $\phi$, such that the overall free energy $\mathcal{F} = \mathcal{F}_\mathrm{B} +  \frac{1}{2} \phi^\dagger Q \phi $ \cite{modelAB}. For systems of our interest, the diffusive dynamics, controlled by $\mu_\mathrm{B}$, drives conservative phase separation. Then the simplest choices for $\mu_\mathrm{B, A}$ (lowest order in $\phi$ and $\nabla$) that has an equilibrium subspace are, 
\begin{equation} 
\eqalign{  
\mu_\mathrm{B} &= c - \alpha \phi + \beta \phi^3 - \kappa \nabla^2 \phi \\
\mu_\mathrm{A} &= c + \alpha' \phi + \beta' \phi^3 
} 
\label{eq:mu_ab}
\end{equation} 
where $\alpha, \beta, \kappa, \alpha', \beta'$ are positive constants and $c$ can take either sign (the constant terms can be taken to be the same w.l.o.g.\ as the constant term in $\mu_\mathrm{B}$ has no effect on the dynamics) . Observe that when $\beta = \beta'$, $\mu_\mathrm{B} - \mu_\mathrm{A} = (\alpha + \alpha') \phi - \kappa \nabla^2 \phi $, which is linear in $\phi$ with self-adjoint linear operator $Q(\boldsymbol{q}) =K(\bq) ^{-1} \left [  (\alpha + \alpha') - \kappa | \bq | ^2  \right ] $. This gives us a continuous parameter $\beta'$ that we can tune to bring the system in and out of the equilibrium subspace without dramatically changing the phenomena exhibited. On its own, the Model B sector favours conservative bulk phase separation with the double-well free energy $\mathcal{F}_\mathrm{B} =  \int \mathrm{d} \bx \left [ - \frac{1}{2}\alpha \phi^2 + \frac{1}{4}\beta \phi^4 + \frac{1}{2}\kappa | \bnabla \phi |^2 \right ]$, whereas the Model A sector describes non-conservative relaxation towards some fixed target density $\phi_\mathrm{t}$. In our previous paper \cite{modelAB}, we surveyed the parameter space of this class of models and found two stable stationary solutions: uniform solution and arrested phase separation. The uniform state is observed for small $M_\mathrm{A}$, where the local nonconservative relaxations overcome the phase separating diffusive dynamics, fixing the density at the target density $\phi_\mathrm{t}$ of the reactions. After crossing over some critical value of $M_\mathrm{A}$, the conservative sector dominates and finite domains of alternating phases are formed, as shown in Fig.\ \ref{fig:steadystate}. The length scale of the pattern is determined by the balance of the reactions in the two phases and the steady state current transporting matter across the interface. Interestingly, the information on these macroscopic currents is not available unless we are able to track the Model B and Model A dynamics separately, i.e.\ distinguishing $\partial_t \phi_\mathrm{A}$ from $\partial_t \phi_\mathrm{B}$ in equation (\ref{eq:phi_dot_ab}). This has profound consequences for the entropy production rate as we will see later. 

\section{Entropy production rate for $(\phi, \partial_t \phi)$ } 
\label{phi_dot_epr} 
For general scalar Langevin systems, a natural choice is to track the evolving field via $(\phi, \partial_t \phi)$ trajectories \cite{nardini2017entropy}. In this section, we work towards a general expression for the instantaneous entropy production rate (EPR) and show that it is non-negative for every $\phi$-configuration in steady state. In addition, we find a non-negative local decomposition of the EPR and discuss its physical interpretations with various examples. Lastly we apply the method to Model AB and, with that as a case study, demonstrate that the EPR can be calculated to lowest order in $\epsilon$ for any Langevin systems. 

Recall from section (\ref{entropy_production}) that the entropy production can be split into an internal part and an external part. We will calculate the external entropy production first, which is the difference between the action of the forward and the backward path as shown in equation (\ref{eq:int_ext}). The action for the $(\phi(x, t), \partial_t \phi(x, t))_{t \in [0, \tau]}$ trajectories of a general Langevin system (see equation (\ref{eq:langevin})) is an Onsager-Machlup functional \cite{OM},
\begin{equation}
\mathbb{A}[\phi, \partial_t \phi ]  = \frac{1}{4} \int_0^\tau \mathrm{d} t \left [ (\partial_t \phi - F )^\dagger K^{-1} (\partial_t \phi - F) + \mathcal{G}[\phi(t)] \right ] 
\end{equation}
where $\mathcal{G}[\phi]$ is a piece specific to Stratonovich discretisation that is only a function of $\phi$ \cite{nardini2017entropy}.  In our vector notations, $\mathcal{G}[\phi] = 2 \epsilon \, \mathrm{Tr}[ \delta_\phi F]$ (here $\delta_\phi F$ is a matrix as $F$ and $\phi$ are both vectors), which in real space is represented by the integral $2 \epsilon \int \mathrm{d} \bx \frac{\delta F(\bx)}{\delta \phi(\bx)}$ \cite{lau}. The reversed $\phi$-path is the direct time reversal of the forward path $\phi^\mathrm{R}(t) = \phi(\tau- t)$, and we differentiate $\phi^\mathrm{R}$ with respect time to get $\partial_t \phi^\mathrm{R}(t) =  - \partial_s \phi (s)|_{s = \tau - t}$. Hence the action for the reversed trajectories is, 
\begin{equation}
\mathbb{A}[\phi^\mathrm{R}, \partial_t \phi^\mathrm{R} ] = \frac{1}{4} \int_0^\tau \mathrm{d} t  \left [ (- \partial_t \phi - F )^\dagger K^{-1} (- \partial_t \phi - F) + \mathcal{G}[\phi]\right ] 
\end{equation}
Note that the additional piece $\mathcal{G}[\phi]$ remains the same, as it does not depend on $\partial_t \phi$. Collecting the two results, the external entropy production and its time derivative (the external EPR) are 
\begin{equation}
\eqalign{  
 \Delta S_\mathrm{Ext} &= - \mathbb{A}[\phi, \partial_t \phi ] + \mathbb{A}[\phi^\mathrm{R}, \partial_t \phi^\mathrm{R} ] =
  \frac{1}{\epsilon} \int_0^\tau \mathrm{d} t \, \partial_t \phi^\dagger K^{-1} F(\phi) \\
\epsilon \dot{S}_\mathrm{Ext}(t) &=  \partial_t \phi^\dagger K^{-1} F(\phi) 
} 
\end{equation}
Next, we take the time derivative of the internal entropy production, which will later be added to $\dot{S}_\mathrm{Ext}(t)$ to give the total EPR. As the starting and end points of the trajectories do not depend on $\partial_t \phi$, the internal entropy production is only a function of $\phi$: $\epsilon \Delta S_\mathrm{Int} = \mathcal{V}[\phi(\tau), \tau]- \mathcal{V}[\phi(0), 0]$. Differentiating with respect to $\tau$ along the trajectories and relabelling $\tau$ to $t$ \cite{seifert}, 
\begin{equation}
\epsilon \dot{S}_\mathrm{Int}(t) = \partial_t \phi^\dagger \frac{\delta \mathcal{V} }{\delta \phi } + \partial_t \mathcal{V} 
\label{eq:s_int} 
\end{equation}
The first term is the change in quasipotential as a result of moving along the trajectory whereas the second comes from the intrinsic time dependence of the probability distribution. Combining the internal and external terms, we obtain (see also \cite{niggemann2020}), 
\begin{equation}
\epsilon \dot{S} = \partial_t \mathcal{V} +   \partial_t \phi^\dagger K^{-1} \left ( F(\phi) + K \frac{\delta \mathcal{V} }{\delta \phi} \right ) \equiv \partial_t \mathcal{V} + \partial_t \phi^\dagger K^{-1} F_\mathrm{a}(\phi) 
\end{equation}
where we have defined a new variable $F_\mathrm{a}$ and the subscript ``a'' denotes that it is the antisymmetric component of the dynamics, as will be explained in the next section. In Stratonovich discretisation, the conditional expectation of $\partial_t \phi$ given the field configuration $\phi$ at time $t$ is $\expval{ \partial_t \phi | \phi, t  }= F_\mathrm{a}(\phi)$ (see appendix \ref{ap:exp} for the precise definition of the conditional expectation and details of the calculation) \cite{seifert, seifert2005entropy}. Averaging over $\partial_t \phi$ of the trajectory, we find that for each state $\phi$ (rather than trajectory $\phi(t)_{t \in [0, \tau] }$), the instantaneous entropy production rate is, 
\begin{equation}
\epsilon \dot{S}[\phi] =\partial_t \mathcal{V} + F_\mathrm{a}^\dagger K^{-1} F_\mathrm{a} 
\label{eq:epr}
\end{equation}

There are two important observations: (1) If we take the ensemble average over the probability distribution $\mathcal{P}(t)$, the first term becomes $\int \mathcal{D}\phi \, \partial_t \mathcal{P}[\phi, t] $ which sums to zero by the conservation of probability \cite{seifert, seifert2005entropy}. Thus the ensemble entropy production rate is always non-negative, consistent with the Second Law.  (2) In steady state $\partial_t \mathcal{V} = 0$, as a result, the instantaneous EPR $\dot{S}_\mathrm{ss}$ is non-negative for any field configuration, 
\begin{equation}
\epsilon \dot{S}_\mathrm{ss} = F_\mathrm{a}^\dagger K^{-1} F_\mathrm{a} 
\label{eq:steadystate}
\end{equation}
Note that this does not imply that trajectories with negative entropy production do not exist, as the non-negative quantity $\dot{S}[\phi]$ is the entropy production associated with each field configuration, which we obtained by averaging over  $\expval{ \partial_t \phi | \phi }$.

The non-negativity of the steady state EPR can be proved by further decomposing $\dot{S}_\mathrm{ss}$ as a sum of non-negative elements in a basis-independent way: $\dot{S}_\mathrm{ss} = \epsilon^{-1} \sum_{i} | Y_i |^2 =  Y^\dagger Y$ for $| Y | = | \sigma^{-1} F_\mathrm{a}|$ (only the magnitude of $Y$ is important but not its phase or direction, e.g.\  $Y = \pm \sigma^{-1} F_\mathrm{a}$ gives the same decomposition in any basis). In particular, this gives a non-negative spatial decomposition $\dot{s}(\bx) = \epsilon^{-1} | Y(\bx) |^2$ such that $\dot{S}_\mathrm{ss} = \int \mathrm{d} \bx \dot{s}(\bx)$ and $|Y|$ in real space plays an analogous role as the antisymmetric current in Macroscopic Fluctuation Theory \cite{mft} as we will see later. However, we would like to emphasise that it is not a \textit{unique} local decomposition (e.g. $\epsilon^{-1} F_\mathrm{a}(\bx) (K^{-1} F_\mathrm{a})(\bx)$ gives the same result upon spatial integration) and \textit{a priori} there is no reason to choose one over the other. The choice of $\dot{s}(\bx)$ is singled out by the fact that it guarantees non-negative local decomposition.   

\subsection{The antisymmetric component $F_\mathrm{a}$} 
\label{f_a} 

Roughly speaking, $F_\mathrm{a}$ ($= F + K \frac{\delta \mathcal{V}}{\delta \phi}$) is the time antisymmetric part of the deterministic evolution; the rigorous version of this statement requires the introduction of the adjoint dynamics. Define the adjoint Fokker-Planck equation such that the time evolution of the probability under the adjoint Fokker-Planck is the same as the forward probability evolution run backwards. Mathematically, this means $\mathcal{P}^\mathrm{R}( \tau - t) = \mathcal{P}(t)$, where $\mathcal{P}^\mathrm{R}$ is the solution of the adjoint Fokker-Planck equation. Following \cite{mft}, assume that the adjoint system is also of Langevin type and has the same noise kernel, we want to find the deterministic adjoint dynamics $F^\mathrm{R}$. 

Starting with the Fokker-Planck equation for the forward probability distribution $\mathcal{P}[\phi, t]$, omitting the dependence on $\phi$ to ease notation, 
\begin{equation}
\partial_t \mathcal{P}(t) = \delta_\phi^\dagger ( F  - \epsilon  K \delta_{\phi}  )  \mathcal{P}(t) 
\label{eq:fokkerplanck}
\end{equation} 
where $\delta_\phi$ denotes functional derivative with respect to the $\phi$-field, which behaves like a vector. On the other hand, the adjoint Fokker-Planck equation is
\begin{equation}
\partial_{t'} \mathcal{P}^\mathrm{R}(t')  = \delta_{\phi}^\dagger  \left (  F^\mathrm{R} - \epsilon  K  \delta_{\phi} \right ) \mathcal{P}^\mathrm{R}(t')  
\end{equation} 
Enforcing the condition that $ \mathcal{P}^\mathrm{R}( \tau - t) = \mathcal{P}(t)$, we must have $t' = \tau - t$ and, by the chain rule, $\partial_{t'} P^\mathrm{R} (t') = - \partial_t \mathcal{P}^\mathrm{R}(\tau - t) = - \partial_t \mathcal{P}(t)$. Substituting into the adjoint Fokker-Planck equation and rearranging the terms into a drift term and noise term, we obtain
\begin{equation} 
\partial_t \mathcal{P}(t)  = \delta_{\phi}^\dagger  [ ( 2 \epsilon K \delta_{\phi} \log( \mathcal{P}(t) ) - F^\mathrm{R} ) \mathcal{P}(t)   - \epsilon K \delta_{\phi}  \mathcal{P}(t)  ] 
\end{equation}
Comparing the above with equation (\ref{eq:fokkerplanck}), we can see that the noise terms are the same and the equations will be identical if the drift terms are matched. Recall that $\mathcal{P} \propto \exp( - \epsilon^{-1} \mathcal{V} )$, we obtain an expression for $F^\mathrm{R}$ in terms of the forward dynamics, 
\begin{equation}
F^\mathrm{R}(\phi, t) = -  2K \delta_\phi \mathcal{V}[\phi, t]  - F(\phi)
\end{equation}
This naturally leads to the definitions of the symmetric and antisymmetric components $F_\mathrm{s, a} \equiv (F \pm F^\mathrm{R})/2$, 
\begin{equation}
F_\mathrm{s}  = - K \frac{\delta \mathcal{V}}{\delta \phi}, \quad F_\mathrm{a} =  F +  K \frac{\delta \mathcal{V} }{\delta \phi} 
\label{eq:f_a}
\end{equation}
Note that the forward Fokker-Planck equation needs to be solved to obtain the adjoint dynamics as it requires the value of $\mathcal{V}(t)$ at every instant in time. However, the situation simplifies in steady state where $\mathcal{V}$ remains invariant. Then $F_\mathrm{s}$ describes the descent to the minimum of the quasipotential, whereas $F_\mathrm{a}$ characterises the excess driving that maintains the system away from equilibrium. It is not surprising that the entropy production in steady state is only a function of $F_\mathrm{a}$ -- both the entropy production and $F_\mathrm{a}$ quantify the amount of time reversal symmetry breaking in the system and vanish in equilibrium. 

Our definitions of $F_\mathrm{s, a}$ are similar to the symmetric and antisymmetric currents $\boldsymbol{J}_\mathrm{s, a}$ in Macroscopic Fluctuation Theory (MFT) \cite{mft}. Both rely on the time reversal of the dynamics, but our derivation seeks the time reversal of the probability evolution $\mathcal{P}(t)$, whereas the adjoint dynamics in MFT  is defined such that the backward path in the adjoint dynamics has the same probability as the forward path in the forward dynamics. We refer to \ref{ap:mft} for the precise formulae for $\boldsymbol{J}_\mathrm{s, a}$ and further discussions on when our adjoint dynamics coincide with that in MFT. 

\subsection{Entropy production in equilibrium}  
\label{eq_entropy}
When the principle of detailed balance holds, the probability of observing a path is the same as the probability of observing the reversed path: $\mathbb{R}[\{\boldsymbol{X}\} ] = \mathbb{R}[\{\boldsymbol{X}^\mathrm{R} \} ]$ \cite{mukamel}. So we expect the total entropy production $\Delta S$ of any trajectory to vanish. In addition, the presence of time reversal symmetry means that all equilibrium systems are self-adjoint. Therefore we expect $F  = F^\mathrm{R}$, $F_\mathrm{a} = 0$ and the instantaneous entropy production rate $\dot{S}[\phi]$ to be zero for all field configurations $\phi$. 

This can be demonstrated more concretely for the relaxational models introduced in section \ref{scalar_langevin}. Recall that for this class of systems, $F = - K \frac{\delta \mathcal{F}}{\delta \phi}$ and the stationary measure is the Boltzmann distribution: $\mathcal{V}[\phi] = \mathcal{F} [\phi]$. Thus, for a $(\phi, \partial_t \phi)_{t \in [0, \tau]}$ trajectory, the internal and external entropy productions in steady state (equilibrium) are, 
\begin{equation}
\eqalign{  
\fl \quad \Delta S_\mathrm{Int} &= \frac{1}{\epsilon} \left (  \mathcal{F}[\phi(\tau)] - \mathcal{F}[\phi(0)] \right ) \\
\fl \quad \Delta S_\mathrm{Ext} &= -  \frac{1}{\epsilon} \int \mathrm{d} t \, \partial_t \phi^\dagger \mu
=  - \frac{1}{\epsilon} \int \mathrm{d} t \, \partial_t \phi^\dagger \frac{\delta \mathcal{F}}{\delta \phi}
= - \frac{1}{\epsilon} \left (  \mathcal{F}[\phi(\tau)] - \mathcal{F}[\phi(0)] \right ) 
} 
\end{equation}
Observe that the two parts cancel out exactly. The total entropy production $\Delta S$ vanishes for each trajectory -- the entropy has merely been transferred from the system to the surroundings but the overall value remains the same. 

As a consistency check, we can also compute the instantaneous entropy production rate $\dot{S}$ via equation (\ref{eq:epr}). Recall from equation (\ref{eq:f_a}) that $F_\mathrm{a} =  F +  K \frac{\delta \mathcal{V} }{\delta \phi}$, which is identically zero in equilibrium since $\mathcal{V}_\mathrm{ss} = \mathcal{F}$. The steady state EPR $\dot{S}_\mathrm{ss}$ and its local decomposition $\dot{s}$ are only functions of $F_\mathrm{a}$ but not $F_\mathrm{s}$, hence must both vanish, as expected of time-symmetric dynamics. 

\subsection{Models with mass conservation} 
\label{conserved}
Having checked that the entropy production indeed vanishes in equilibrium, we proceed to probe the class of non-equilibrium diffusive systems introduced in section \ref{relax} that breaks time reversal symmetry in a `minimal' way by adding a driving term to Model B. We will compute the steady state entropy production rate $\dot{S}_\mathrm{ss}$ and its non-negative local decomposition $\dot{s}(\bx)$, followed by a discussion of their physical interpretations. For convenience, throughout this section we will omit the subscript `B' as we only discuss mass-conserving systems. 

Recall from section \ref{relax} that the noise kernel $K$ for mass-conserving systems is $- M \nabla^2$, from which we can straight forwardly deduce the antisymmetric component $F_\mathrm{a}$ and a formal expression for $\dot{S}_\mathrm{ss}$ 
\begin{equation}
\eqalign{  
F_\mathrm{a} &= F + K \delta_\phi \mathcal{V}_\mathrm{ss} = M \bnabla \cdot \left [ \bnabla \left ( \mu - \delta_\phi \mathcal{V}_\mathrm{ss} \right ) - \boldsymbol{E} \right ]  \\
\dot{S}_\mathrm{ss} &=\frac{1}{\epsilon M} \int \mathrm{d}\bx \, F_\mathrm{a}(\bx) (\nabla^{-2} F_\mathrm{a})(\bx) 
} 
\end{equation}
The inverse of the Laplacian operator, denoted as $\nabla^{-2}$, is well defined up to a constant once the boundary conditions are specified. In fact, we can infer that the constant piece must be zero because both $\mathcal{V}$ and $\mathbb{A}$ have no contribution from the $\bq = 0$ mode due to mass conservation, meaning that the entropy production rate cannot have a zero-mode contribution either.  

The integral in the expression for $\dot{S}_\mathrm{ss}$ gives one spatial decomposition, but that is not the non-negative local EPR $\dot{s}(\bx) $. Recall that $\dot{s}(\bx) = \epsilon^{-1} |Y(\bx)|^2$, where $| Y | = | \sigma^{-1} F_\mathrm{a} |$ and $\sigma(\bq) = - i \sqrt{M} | \bq |$. The subtleties with $\sigma(\bq = 0) = 0$ can again be mitigated by realising that there cannot be a $\bq = 0$ mode in the entropy production rate. For mass-conserving systems, as alluded to in section \ref{f_a} and discussed in \ref{ap:mft}, $F_\mathrm{a}$ is closely related to the antisymmetric current $\boldsymbol{J}_\mathrm{a} = - \bnabla \left ( \mu - \delta_\phi \mathcal{V}_\mathrm{ss} \right ) + \boldsymbol{E}$ in Macroscopic Fluctuation Theory. In fact, $F_\mathrm{a} = - \div \boldsymbol{J}_\mathrm{a}$, which in Fourier space translates to $F_\mathrm{a} (\bq) = i \bq \boldsymbol{\cdot} \boldsymbol{J}_\mathrm{a} (\bq)$, implying that we can choose $Y(\boldsymbol{q}) = M^{-1/2} \hat{\bq}  \boldsymbol{\cdot} \boldsymbol{J}_\mathrm{a}$, where $\hat{\bq} = \bq/ | \bq |$, with the additional condition that $Y(\boldsymbol{q}=0)= 0$.  The physical interpretation of this becomes clearer if we decompose $\boldsymbol{J}_\mathrm{a}$ into a pure gradient piece and a pure curl piece,
\begin{equation}
\boldsymbol{J}_\mathrm{a} = \bnabla \Phi +  \curl A
\end{equation}
where $\Phi$ is a scalar field and $A$ is a vector field. This is the Helmholtz decomposition, which can always be performed for any vector field: $\Phi = \nabla^{-2} \div \boldsymbol{J}_\mathrm{a}$, where the inverse Laplacian is well defined as described above. Let the pure gradient part be $\widetilde{\boldsymbol{J}}_\mathrm{a} = \bnabla \Phi$. Since $\widetilde{\boldsymbol{J}}_\mathrm{a}(\bq)= - i \boldsymbol{q} \Phi(\bq)$, $\widetilde{\boldsymbol{J}}_\mathrm{a}$ is in the direction of $\boldsymbol{q}$ while $(\curl \boldsymbol{A})(\bq)= - i \bq \times \boldsymbol{A}(\bq)$ is perpendicular to $\bq$. Therefore
\begin{equation}
| Y (\bq) |  = M^{-1/2} | \hat{\boldsymbol{q}} \boldsymbol{\cdot} \boldsymbol{J}_\mathrm{a} (\bq) | 
= M^{-1/2} | \widetilde{\boldsymbol{J}}_\mathrm{a} (\bq) | 
\end{equation}
Fourier transform back to real space, 
\begin{equation}
\dot{s}(\bx ) = \epsilon^{-1} | Y(\bx) |^2 = (\epsilon M)^{-1} \widetilde{\boldsymbol{J}}_\mathrm{a} (\bx) \cdot
\widetilde{\boldsymbol{J}}_\mathrm{a} (\bx)
\label{eq:modelb_entropy}
\end{equation}
In other words, any curl piece (including any constant as this cannot be written as a gradient) in the current gives no contribution to the entropy production rate, which makes sense as the non-gradient part of the current has no effect on the $(\phi, \partial_t \phi)$ trajectories either. 

Recall the simple exactly solvable example of particles driven around a 1D ring by a constant force $\boldsymbol{\gamma}$, as discussed in section \ref{scalar_langevin}. In this case, the steady state probabilities are independent of the driving (i.e.\ $\mathcal{V}_\mathrm{ss} = \mathcal{F}$), leading to a clear decomposition of the symmetric and antisymmetric dynamics 
\begin{equation}
F_\mathrm{s} = M \partial_x^2 \mu, \quad F_\mathrm{a} = - M \gamma \partial_x \phi
\end{equation}
The symmetric part controls the descent down the free energy gradient while the antisymmetric part drives the system around the ring. In 1D, the only curl contribution in $J_\mathrm{a}$ is the constant piece, which we need to subtract off: $\widetilde{J}_\mathrm{a}(x) = M \gamma (\phi(x) - \bar{\phi})$, where $\bar{\phi}$ is the mean density. Therefore the local entropy production is 
\begin{equation}
\dot{s}(x) = \epsilon^{-1} M |\boldsymbol{\gamma}|^2  \left (\phi(x) - \bar{\phi} \right )^2 
\end{equation}
For more complicated driving, such as Active Model B+, the quasipotential $\mathcal{V}$ is often unknown so it is usually not possible compute the entropy production exactly. 

\subsection{Model AB} 
Similarly, for Model AB, there is no known solution for the full stochastic dynamics. Nevertheless, one can still write down an expression for the steady state entropy production, assuming that $\mathcal{V}_\mathrm{ss}$ can at least be approximated. Recall the definition of Model AB from equation (\ref{eq:modelab}): $\partial_t \phi = M_\mathrm{B} \nabla^2 \mu_\mathrm{B} - M_\mathrm{A} \mu_\mathrm{A} + \sqrt{2 \epsilon} \sigma \Lambda$ where $\sigma(\boldsymbol{q}) = \sqrt{M_\mathrm{A}} - i | \bq | \sqrt{M_\mathrm{B}}$. The antisymmetric component in steady state is   
\begin{equation}
F_\mathrm{a} = M_\mathrm{B} \nabla^2 \left (  \mu_\mathrm{B}  - \frac{\delta \mathcal{V}_\mathrm{ss}}{\delta \phi} \right ) - M_\mathrm{A} \left (  \mu_\mathrm{A} - \frac{\delta \mathcal{V}_\mathrm{ss}}{\delta \phi} \right ) 
\end{equation}
where as before $\mathcal{V}_\mathrm{ss}$ is the (unknown) steady state quasipotential. Recall that the local decomposition $\dot{s}(\bx) = \epsilon^{-1} Y^2$ where $| Y  | = | \sigma^{-1} F_\mathrm{a} |$. Since $\sigma$ is invertible and diagonal in Fourier space, given $\mathcal{V}_\mathrm{ss}$, $Y$ can be computed independently for each mode:  $| Y(\boldsymbol{q}) | = |F_\mathrm{a}(\boldsymbol{q})/\sigma(\boldsymbol{q})|$. It is possible to then Fourier transform back to real space to obtain an expression for $\dot{s}(\bx)$, but we will not write it out explicitly here as it's both complicated and not particularly enlightening. 

\subsection{Small noise expansion} \label{small_noise} 

As we have shown with examples, in general it is extremely rare that one can find an exact analytical expressions for $\dot{S}_\mathrm{ss}$  or its local decomposition $\dot{s}(\bx) $, as they require the knowledge of $F_\mathrm{a}$ which in turn depends on an exact solution for the steady state probability distribution (recall that $F_\mathrm{a} = F + K \frac{\delta \mathcal{V}}{\delta \phi}$ and $\mathcal{P}_\mathrm{ss} = \exp ( - \epsilon^{-1} \mathcal{V}_\mathrm{ss})$). However, when the noise strength $\epsilon$ is small, the stationary distribution is approximately a Gaussian distribution around the deterministic steady state as we will see shortly\footnote{In cases where the quasipotential has multiple minima, the approximation is valid for the time window before the escape time of the local minimum. }. As a result, the steady state EPR can be approximated in the small $\epsilon$ limit. 

The small noise approximation is a standard method for stochastic processes: first expand the scalar field $\phi = \phi_0 + \sqrt{\epsilon} \phi_1 + O(\epsilon)$, then substitute into equation (\ref{eq:langevin}) and equate the terms to order $\epsilon^0$ and $\epsilon^{1/2}$ separately \cite{gardiner, vankampen}, 
\begin{equation}
\eqalign{  
\partial_t \phi_0 &= F (\phi_0) \\
\partial_t \phi_1 & = A (\phi_0)  \phi_1 + \sqrt{2} \sigma  \Lambda 
} 
\end{equation}
Here $A(\phi_0) $ is the Jacobian ``matrix'' defined as $A_{ij}(\phi_0) \equiv \delta_{\phi_j} F_i |_{\phi = \phi_0}$. Note that $A$ is a function of $\phi_0$ but not $\phi_1$. The zeroth order field $\phi_0$ captures the deterministic evolution while the $\phi_1$ equation corresponds to the Gaussian fluctuation around the deterministic trajectory. Once $\phi_0$ has reached its stationary value $\phi_0^\mathrm{ss}$, the Jacobian $A$ must be negative definite (all eigenvalues negative)\footnote{Modulo any Goldstone mode -- see \ref{ap:symmetry}}. This means $\phi_1$ decays exponentially towards a steady state with mean value zero. The steady state correlation of $\phi_1$,  defined as $C = \expval{ \phi_1 \phi_1^\dagger }$, can be calculated using the Lyapunov equation \cite{gardiner, vankampen}, 
\begin{equation}
AC + C A^\dagger = - 2 K 
\end{equation}
Once $C$ is known, the steady state quasipotential $\mathcal{V}_\mathrm{ss}$ can also be obtained to lowest order in $\epsilon$. As $\phi_0$ is entirely deterministic, the quasipotential is only a function of the fluctuating field $\phi_1$, 
\begin{equation}
\mathcal{V}_\mathrm{ss}[ \phi_1 ] =   \frac{\epsilon}{2} \phi_1 ^\dagger G \phi_1 + O(\epsilon^{3/2})
\label{eq:quasipotential}
\end{equation}
where $G$ is the inverse of $C$. This holds provided that the system has no continuous symmetry (e.g. translation symmetry if periodic boundary conditions are used); otherwise the additional degree of symmetry has to be explicitly projected out as shown in \ref{ap:symmetry} so that $G = C^{-1}$ in the subspace without the Goldstone mode. Now we have all the pieces to calculate the time-antisymmetric component $F_\mathrm{a}$ and $Y = \sigma^{-1} F_\mathrm{a}$, 
\begin{equation}
\eqalign{  
F_\mathrm{a} &= F(\phi) + K \frac{\delta \mathcal{V}_\mathrm{ss}}{\delta \phi} 
= \sqrt{\epsilon} ( A + K G) \phi_1 + O(\epsilon) \\
Y & = \sqrt{\epsilon} \left ( \sigma^{-1} A + \sigma^\dagger G \right ) \phi_1 + O(\epsilon) 
} 
\end{equation}
For convenience, define the matrix $E = \sigma^{-1} A + \sigma^\dagger G$ such that $Y = \sqrt{\epsilon} E \phi_1 + O(\epsilon)$. Notice that both $F_\mathrm{a}$ and $Y$ are $O(\sqrt{\epsilon})$, implying that they vanish at the deterministic order and only depend on the fluctuations. Finally, we can write down the local entropy production rate for a steady state configuration $\phi = \phi_0 + \sqrt{\epsilon} \phi_1 + O(\epsilon) $, 
\begin{equation}
\fl \dot{s}(\bx)[\phi_1]  = \epsilon^{-1} Y(\bx)^2 
= \int \mathrm{d} \boldsymbol{y} \mathrm{d} \boldsymbol{z} E(\bx, \boldsymbol{y}) \phi_1(\boldsymbol{y}) E(\bx, \boldsymbol{z}) \phi_1(\boldsymbol{z}) + h.o.t.
\end{equation}
where $h.o.t.$ stands for `higher order terms'. We can proceed to average over the steady state distribution for $\phi_1$ to obtain the ensemble EPR, 
\begin{equation}
\eqalign{  
\expval { \dot{s}(\bx) } &= \int \mathrm{d} \boldsymbol{y} \mathrm{d} \boldsymbol{z} E(\bx, \boldsymbol{y}) C(\boldsymbol{y}, \boldsymbol{z} ) E^\dagger (\boldsymbol{z}, \bx)  + h.o.t. \\
&= \mathrm{Diag} [ E C E^\dagger ](\bx) + h.o.t.
} 
\label{eq:local_phi_EPR}
\end{equation}
where ``Diag'' denotes the diagonal of a matrix. It is worth noting that we also get $\expval{\dot{S}_\mathrm{ss}[\phi]}$ for free: $\expval{\dot{S}_\mathrm{ss}} = \int \mathrm{d}\bx \expval{ \dot{s}(\bx) } = \mathrm{Tr}( E C E^\dagger) + h.o.t. $ Thus far, we have arrived at a general expression for the lowest order noise expansion for the steady state EPR of any scalar Langevin system. The only knowledge required for this calculation is the deterministic steady state solution, which is both more analytically tractable and less numerically expensive than solving the stochastic dynamics. 

Interestingly, we can now identify a sufficient condition for the entropy production to vanish at the leading order in $\epsilon$: when $A, K$ are simultaneously diagonalisable. Let the shared eigenstates be $\{v_i\}$ and the corresponding eigenvalues for $A$ and $K$ be $\{a_i\}$ and$\{k_i\}$ respectively\footnote{If there are symmetries or conservation laws, we only work in the relevant physical subspace. See \ref{ap:symmetry} and section \ref{conserved}}. In this shared eigenspace spanned by $\{v_i\}$, the Lyapunov equation simplifies to $C_{ij} a_j^* + C_{ij} a_i = - 2 k_i \delta_{ij} $ (no summation convention). Since $A$ is negative definite, implying that $\{a_i\}$ are all real negative numbers, $C_{ij} = - 2 k_i \delta_{ij}/ (a_i + a_j)$, from which we conclude that all off-diagonal elements of $C$ vanish whereas the diagonal elements are known exactly: $C_{ii} = - k_i/ a_i$. Substituting into the expression for $E$, we see that $E$ is also diagonal in the eigenspace and its eigenvalues $\{e_i\}$ are, 
\begin{equation} 
e_i = \sigma_i^{-1} a_i +\sigma_i^* C_{ii}^{-1} = \sigma_i^{-1} a_i - \sigma_i^* a_i/  k_i = 0
\end{equation}
where the definition of $\sigma$ has been used to deduce $\sigma_i \sigma_i^* = k_i$. 
As a result, for any steady state field fluctuation $\phi_1$ around the stationary solution $\phi_0^\mathrm{ss}$, we have $Y = \sqrt{\epsilon} E \phi_1 + O(\epsilon) = O(\epsilon)$ and the local EPR $\dot{s}(\bx)$ associated with any configuration is at least of order $\epsilon$. 

To sum up, we have arrived at a general result for the ensemble average of steady state EPR $\expval { \dot{s}(\bx)}$ at $O(\epsilon^0)$ that only requires the deterministic solution as an input. Moreover, if the Jacobian $A$ and the noise kernel $K$ can be simultaneously diagonalised, the EPR becomes at least $O(\epsilon)$. 

\subsection{Small noise expansion of Model AB} 
\label{entropy_modelab}
The small noise expansion is especially useful when the steady state of the full stochastic equations cannot be solved analytically. Such is the situation for Model AB away from the special equilibrium subspace described in section \ref{modelAB}. The general scheme is independent of the basis chosen and it turns out to be much simpler in Fourier space. We will therefore perform our algebraic manipulations in Fourier space throughout and only transform back to the real space in the end to calculate $\dot{s}(\bx)$. 

The first step of the calculation is to obtain the Jacobian matrix $A$ from the steady state solution $\phi_0^\mathrm{ss}$, which in Fourier space is explicitly defined as $A(\boldsymbol{q}, \boldsymbol{q}_1) =\delta_{\phi(\boldsymbol{q}_1)} F(\boldsymbol{q}) |_{\phi = \phi_0^\mathrm{ss}}$. Recall from equation (\ref{eq:modelab}, \ref{eq:mu_ab}) that 
\begin{equation}
\eqalign{  
F(\bq) &= - M_\mathrm{A} \mu_\mathrm{A}(\boldsymbol{q}) - M_\mathrm{B}q^2 \mu_\mathrm{B}(\boldsymbol{q}) \\
\mu_\mathrm{A}(\bq) &= c \, \delta_{\bq, 0} + \alpha' \phi(\bq) + \beta' \sum_{\bq_1, \bq_2} \phi(\bq_1) \phi(\bq_2) \phi(\bq - \bq_1 - \bq_2) \\
\mu_\mathrm{B}(\bq) &= - \alpha \phi(\bq) + \kappa q^2 \phi(\bq)  + \beta  \sum_{\bq_1, \bq_2} \phi(\bq_1) \phi(\bq_2) \phi(\bq - \bq_1 - \bq_2) 
} 
\end{equation}
where $\sum_{\bq}$ is a short-hand for integrating over $\mathrm{d}\bq/(2\pi)^d$ in infinite domain and summing over all $\bq$ modes in finite domain. Similarly $\delta_{\bq, \bq'}$ is the appropriate identity matrix in the Fourier space. Performing the functional derivative, 
\begin{equation}
\eqalign{  
\fl A(\bq, \bq_1) =  M_\mathrm{A} \left [ c \, \delta_{\bq, 0} \delta_{\bq_1, 0} + \alpha ' \delta_{\bq, \bq_1} + 3 \beta ' \sum_{\bq_2}  \phi_0^\mathrm{ss}(\bq_2) \phi_0^\mathrm{ss}(\bq - \bq_1- \bq_2) \right ] \\
\fl \qquad \qquad + M_\mathrm{B} q^2 \left [ ( - \alpha +\kappa q^2) \delta_{\bq, \bq_1} + 3 \beta \sum_{\boldsymbol{q}_2} \phi_0^\mathrm{ss}(\boldsymbol{q}_2) \phi_0^\mathrm{ss}(\bq- \bq_1 - \bq_2)   \right ] 
} 
\label{eq:jacobian} 
\end{equation}
We can see that $A$ is in general non-diagonal in Fourier space and it is indeed a function of the deterministic solution $\phi_0^\mathrm{ss}$. As discussed in section \ref{modelAB}, there are typically two stable stationary solutions: the uniform state and arrested phase separation.

\begin{figure}
\begin{subfigure}{0.44\textwidth}
\includegraphics[width=\textwidth]{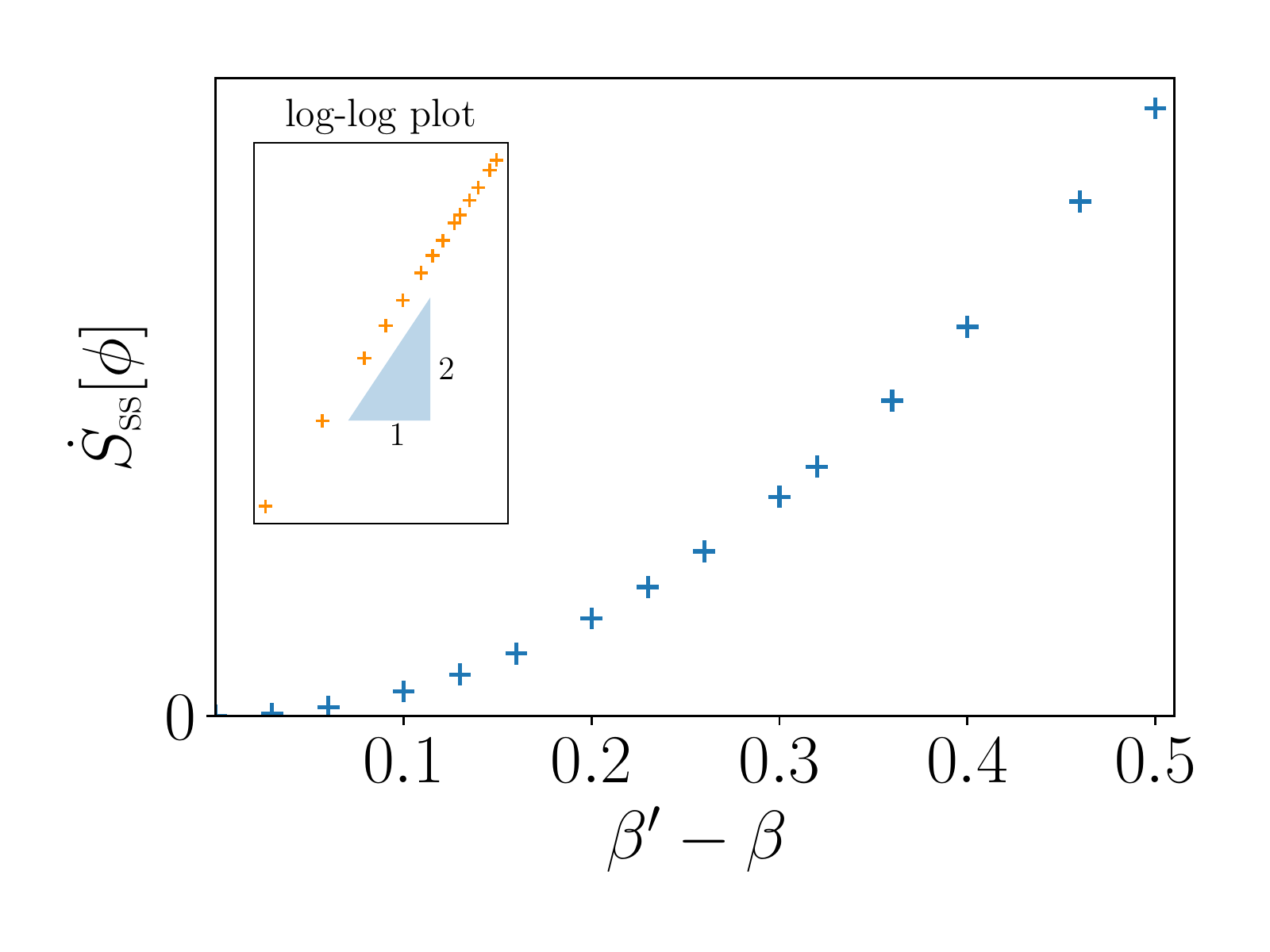}
\caption{}
\label{fig:epr_tot}
\end{subfigure} 
\begin{subfigure}{0.52\textwidth}
\includegraphics[width=\textwidth]{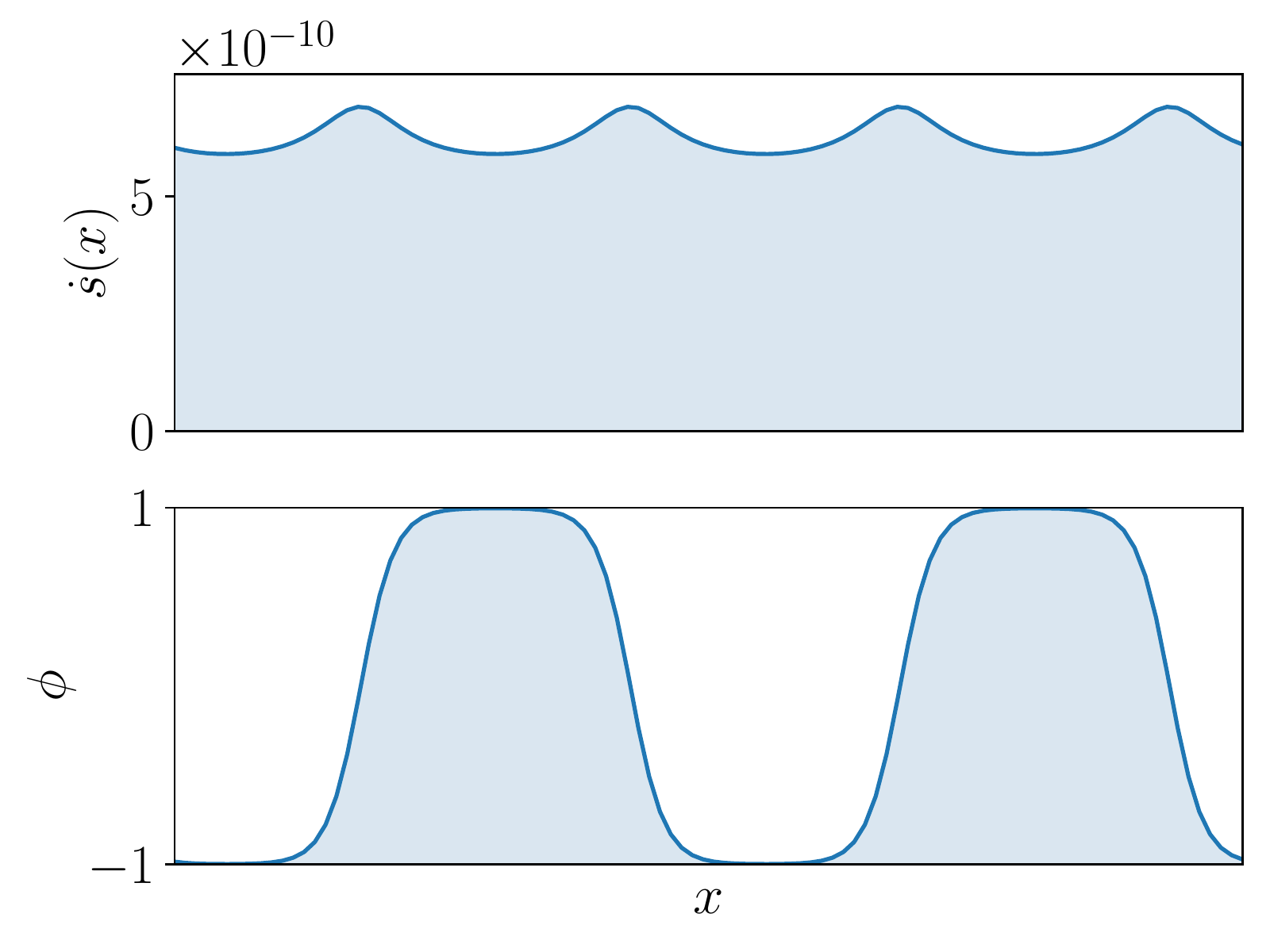}
\caption{}
\end{subfigure} 
\caption{(A) Plot of $\dot{S}_\mathrm{ss}[\phi]$ against $\beta' - \beta$ along with the same data on log-log scale, showing $\dot{S}_\mathrm{ss}[\phi] \propto (\beta' - \beta)^2$ for small $\beta' - \beta$. Observe that the entropy production rate indeed goes to zero as the equilibrium subspace ($\beta' = \beta$) is approached. (B) The upper panel shows the local entropy production rate $\dot{s}(x)$ (in arbitrary units) and the lower panel shows the corresponding $\phi(x)$. The spatial distribution is mostly uniform with small peaks at the interface. The parameter values are $\alpha=\beta=\alpha'=1, \beta' = 1.1, \kappa=5, M_\mathrm{B} = 0.1, M_\mathrm{A} = 5 \times 10^{-6}$.}
\label{fig:epr_phi}
\end{figure}

In the uniform state, $\phi_0^\mathrm{ss}(\bq) = \phi_\mathrm{t} \delta_{\bq, 0}$, where $\phi_\mathrm{t}$ is the target density of the Model A sector. Substituting $\phi_0^\mathrm{ss}$ into the expression for $A$, we can see that $A(\bq, \bq_1)$ is nonzero only if $\bq = \bq_1$. In other words, $A$ is diagonal in Fourier space. Recall that the noise kernel $K$ is also diagonal in Fourier space with diagonal elements $K(\bq) = M_\mathrm{A} + M_\mathrm{B} q^2$. We can conclude that $A, K$ are simultaneously diagonalisable and consequently $\dot{s}(\bx) = O(\epsilon)$. This is echoed by the work of Nardini et al.\ \cite{nardini2017entropy}: they found the entropy production rate of Active Model B to also be $O(\epsilon)$ in the uniform state. In fact, our line of argument directly applies to their model  -- the Jacobian and noise kernel for Active Model B are both diagonal in Fourier space, unless $\phi^0_\mathrm{ss}$ is non-uniform.

On the other hand, in the phase separated state, $A(\bq, \bq_1)$ has nonzero off-diagonal elements and $A, K$ are no longer simultaneously diagonalisable. Consequently, the entropy production rate is of order $\epsilon^0$, again echoing the findings of Nardini et al.\ for the non-homogeneous state of Active Model B \cite{nardini2017entropy}. 

We have explored numerically small values of $\beta' - \beta$ that lead to approximately the same spatial pattern as the equilibrium case ($\beta' = \beta$) to highlight the role of the entropy production rate as a measure of the irreversibility in the system. We refer to section \ref{small_noise} for details of the spatial discretisation method and subtleties associated with finite domains. The final results are shown in Fig.\ \ref{fig:epr_phi}. The total steady state entropy production $\dot{S}_\mathrm{ss}[\phi]$ vanishes in the equilibrium subspace as expected. The spatial decomposition $\dot{s}(x)$ is indeed non-negative everywhere with small peaks at the interfaces, where particles are pumped from the dilute phase to the dense phase. 

Additionally, we found that $\dot{S}_\mathrm{ss}[\phi] \propto (\beta' - \beta)^2$ for the range of $\beta' - \beta$ probed, as shown in Fig.\ \ref{fig:epr_tot}. This can be explained via a perturbative calculation around the equilibrium subspace. Let $\Delta = \beta' - \beta$, and recall from section \ref{modelAB} that the deterministic dynamics $F$ is the sum of the equilibrium part $(- K \delta_\phi \mathcal{F})$ and an order $\Delta$ non-equilibrium piece. Then the Jacobian matrix $A$ shows a similar splitting: $A = - K H + O(\Delta)$, where $H$ is the Hessian of the free energy $\mathcal{F}$, defined as $H_{ij} = \delta_{\phi_i} \delta_{\phi_j} \mathcal{F}$. Note that $H$ is by definition symmetric. The Lyapunov equation can be solved perturbatively yielding the correlation matrix $C = H^{-1} + O(\Delta)$ and its inverse $G = H + O(\Delta)$\footnote{With the usual caveats regarding inverting matrices in the presence of Goldstone modes -- see \ref{ap:symmetry}}. As a result, $E = \sigma^{-1} ( A + K G) = O(\Delta)$ and $\dot{S}_\mathrm{ss} = \mathrm{Tr} ( E CE^\dagger)  = O(\Delta^2)$ as observed. 

So far, we have presented a general formula for the entropy production rate $\dot{S}[\phi]$ of the $(\phi, \partial_t \phi)$ trajectories, as well as a non-negative spatial decomposition $\dot{s}(\bx)$ of its steady state value $\dot{S}_\mathrm{ss}[\phi]$. We demonstrated the computation with a small noise expansion for Model AB, showing that the EPR indeed vanishes as equilibrium is approached and scales as expected with the deviation from equilibrium $(\beta' - \beta)$. We emphasis that all the computations we have performed until now are of the entropy production rate of the $\phi$-trajectories and we do not expect the same conclusions if different information is tracked, as we will show in the next section.

\section{Entropy production rate for $(\phi, \partial_t \phi_\mathrm{A}, \partial_t \phi_\mathrm{B}) $ in Model AB}
\label{phi_dot_a_epr} 

In this section, we explore the consequences of tracking more information in addition to the $\phi$-evolution. A natural choice is to separate the local density change $\partial_t \phi$ into Model A and Model B contributions $\partial_t \phi = \partial_t \phi_\mathrm{A} + \partial_t \phi_\mathrm{B}$, as defined in equation (\ref{eq:phi_dot_ab}). Now we repeat the entire EPR calculation for the new trajectories $(\phi, \partial_t \phi_\mathrm{A}, \partial_t \phi_\mathrm{B})$, including deriving the new expressions for the internal and external entropy productions. 

For the internal entropy production rate $\dot{S}_\mathrm{Int}$, our expression for the $(\partial_t \phi, \phi)$ trajectories can be recycled because the initial condition for the path is only a function of $\phi$ rather than the time derivatives. Substituting $\partial_t \phi_\mathrm{A} + \partial_t \phi_\mathrm{B}$ for $\partial_t \phi$ in equation (\ref{eq:s_int}) yields
\begin{equation}
\epsilon \dot{S}_\mathrm{Int} = (\partial_t \phi_\mathrm{A} + \partial_t \phi_\mathrm{B} ) ^\dagger \frac{\delta \mathcal{V}}{\delta \phi} + \partial_t \mathcal{V}
\end{equation}
The action functional for the new trajectories is more complicated. Applying the standard derivation for the Onsager-Machlup function \cite{OM} to $(\partial_t \phi_\mathrm{A}, \phi)$ and $(\partial_t \phi_\mathrm{B}, \phi)$ in equation (\ref{eq:phi_dot_ab}) separately,
\begin{equation}
\eqalign{  
\fl \mathbb{A}[\phi, \partial_t{\phi}_\mathrm{A}, \partial_t{\phi}_\mathrm{B} ] = \frac{1}{4} \int \mathrm{d}t \, \left ( \left (\partial_t {\phi}_\mathrm{B} + K_\mathrm{B} \mu_\mathrm{B} \right )^\dagger  K_\mathrm{B}^{-1} \left (\partial_t {\phi}_\mathrm{B} + K_\mathrm{B} \mu_\mathrm{B} \right ) + \mathcal{G}_\mathrm{A}[\phi] \right ) \\
+ \frac{1}{4M_\mathrm{A} } \int \mathrm{d}t \, \left (  \left (\partial_t {\phi}_\mathrm{A} + M_\mathrm{A} \mu_\mathrm{A} \right )^
\dagger  \left (\partial_t {\phi}_\mathrm{A} + M_\mathrm{A} \mu_\mathrm{A} \right ) + \mathcal{G}_\mathrm{B}[\phi] \right )
} 
\end{equation}
where $\mathcal{G}_\mathrm{A, B}[\phi]$ are functions of $\phi$ only, as a consequence of the Stratonovich convention for the path integral. One can check that this yields the action for $(\phi, \partial_t \phi)$ path once we change the variable to $\partial_t \phi = \partial_t \phi_\mathrm{A} + \partial_t \phi_\mathrm{B}$ and integrate over one of $(\partial_t \phi_\mathrm{A}, \partial_t \phi_\mathrm{B})$. As before, under time reversal, the time derivatives flip sign: $\partial_t \phi_\mathrm{A, B}^\mathrm{R}(t) = - \partial_s \phi_\mathrm{A, B}(s)|_{s = \tau - t}$. After repeating the steps taken previously (section \ref{phi_dot_epr}), we take the difference between the action of the forward path and that of the reversed path to obtain the external entropy production rate, 
\begin{equation}
\epsilon \dot{S}_\mathrm{Ext} = - \partial_t {\phi}_\mathrm{B}^\dagger \mu_\mathrm{B} - \partial_t {\phi}_\mathrm{A}^\dagger \mu_\mathrm{A} 
\end{equation}
Adding $\dot{S}_\mathrm{Int}$ and  $\dot{S}_\mathrm{Ext}$ gives the total instantaneous entropy production rate, 
\begin{equation}
\fl \quad \epsilon \dot{S} [\phi, \partial_t \phi_\mathrm{A}, \partial_t \phi_\mathrm{B} ] =  \partial_t \mathcal{V} - \partial_t \phi_\mathrm{B}^\dagger \left ( \mu_\mathrm{B} - \frac{\delta \mathcal{V}}{\delta \phi} \right ) - \partial_t \phi_\mathrm{A}^\dagger \left ( \mu_\mathrm{A} -\frac{\delta \mathcal{V}}{\delta \phi} \right ) 
\end{equation}
Once again, we average over the time derivatives $(\partial_t \phi_\mathrm{A}, \partial_t \phi_\mathrm{B})$ using the conditional expectations $\expval{ \partial_t {\phi}_\mathrm{A} | \phi, t} = - M_\mathrm{A} \left ( \mu_\mathrm{A} - \delta_\phi \mathcal{V} \right )$ and $\expval{ \partial_t {\phi}_\mathrm{B} | \phi, t} = - K_\mathrm{B} \left ( \mu_\mathrm{B} -  \delta_\phi \mathcal{V} \right )$. We refer to \ref{ap:exp} for a rather involved calculation of the conditional expectations, but one can check that they indeed add up to the same $\expval{\partial_t {\phi} | \phi, t }$ as before. This yields the total entropy production rate for the configuration $\phi$  
\begin{equation}
\fl \quad \epsilon \dot{S}^\mathrm{\scriptscriptstyle AB}[\phi]  =  \partial_t \mathcal{V} + \left ( \mu_\mathrm{B} -  \delta_\phi \mathcal{V} \right )^\dagger  K_\mathrm{B}  \left ( \mu_\mathrm{B} - \delta_\phi \mathcal{V} \right ) + M_\mathrm{A} \left ( \mu_\mathrm{A} - \delta_\phi \mathcal{V} \right ) ^\dagger  \left ( \mu_\mathrm{A} - \delta_\phi \mathcal{V} \right )
\end{equation}
Notice that $\dot{S}^\mathrm{\scriptscriptstyle AB}$ possesses similar properties as the entropy production rate $\dot{S}[\phi]$ in equation (\ref{eq:epr}): non-negative upon averaging over the stochastic trajectories and non-negative for each field configuration $\phi$ in steady state. Furthermore, we can also decompose the new steady state EPR, denoted as $\dot{S}^\mathrm{AB}_\mathrm{ss}[\phi]$, into local non-negative contributions $\dot{s}^\mathrm{\scriptscriptstyle AB}(\bx)$ such that $\dot{S}^\mathrm{AB}_\mathrm{ss}[\phi] = \int \mathrm{d} \bx \dot{s}^\mathrm{AB}(\bx)$,  
\begin{equation}
\epsilon \dot{s}^\mathrm{\scriptscriptstyle AB}[\bx] =Y_\mathrm{B}(\bx) Y_\mathrm{B}(\bx)  + Y_\mathrm{A}(\bx)  Y_\mathrm{A}(\bx) 
\end{equation}
where $Y_\mathrm{B} = \sigma_\mathrm{B} \left ( \mu_\mathrm{B} - \delta_\phi \mathcal{V}_\mathrm{ss} \right )$and $Y_\mathrm{A}  =  \sqrt{M_\mathrm{A}} \left ( \mu_\mathrm{A}  - \delta_\phi \mathcal{V}_\mathrm{ss} \right )$. Following the arguments made in section (\ref{conserved}), we can see that, if given the quasipotential $\mathcal{V}_\mathrm{ss}$, $Y_\mathrm{B} (\bx)^2= M_\mathrm{B}^{-1/2} |  \tilde{\boldsymbol{J}}_\mathrm{a} (\bx) |^2$, where $\tilde{\boldsymbol{J}}_\mathrm{a} (\bx)$ is the curl-free piece of the diffusive current and equals $\bnabla (\mu_\mathrm{B} - \delta_\phi \mathcal{V}_\mathrm{ss} )$ in this case. Thus we can assign $Y_\mathrm{B}^2$ as the entropy production rate of Model B with quasipotential $\mathcal{V}_\mathrm{ss}$. Similar arguments can be made for $Y_\mathrm{A}$: recall from equation (\ref{eq:modelA,modelB}) that for Model A, $K_\mathrm{A} = M_\mathrm{A}$ and $\sigma_\mathrm{A} = \sqrt{M_\mathrm{A}}$; hence we obtain the antisymmetric component $F^\mathrm{A}_\mathrm{a} =M_\mathrm{A} \left (  \mu_\mathrm{A} - \partial_\phi \mathcal{V}_\mathrm{ss} \right ) $ and the corresponding $Y_\mathrm{A} = \sigma^{-1} F^\mathrm{A}_\mathrm{a}  =  \sqrt{M_\mathrm{A}} \left ( \mu_\mathrm{A}  - \delta_\phi \mathcal{V}_\mathrm{ss} \right )$. Note that the quasipotential $\mathcal{V}_\mathrm{ss}$ is that of the Model AB dynamics, which is qualitatively different from Model A or Model B on its own. 

We would like to emphasize that this clear splitting in terms of Model A and Model B contributions is a consequence of tracking the two sectors separately via $(\partial_t \phi_\mathrm{B}, \partial_t \phi_\mathrm{B})$. Although both $\dot{S}^\mathrm{AB}_\mathrm{ss}[\phi]$ and the previously calculated entropy production rate $\dot{S}_\mathrm{ss}[\phi]$ (equation (\ref{eq:steadystate})) are functionals of the field configuration $\phi$ only, once the time derivatives have been averaged over, they are entirely different quantities and there is no direct path to convert between them. 

\begin{figure}
\begin{subfigure}{0.4\textwidth}
\includegraphics[width=\textwidth]{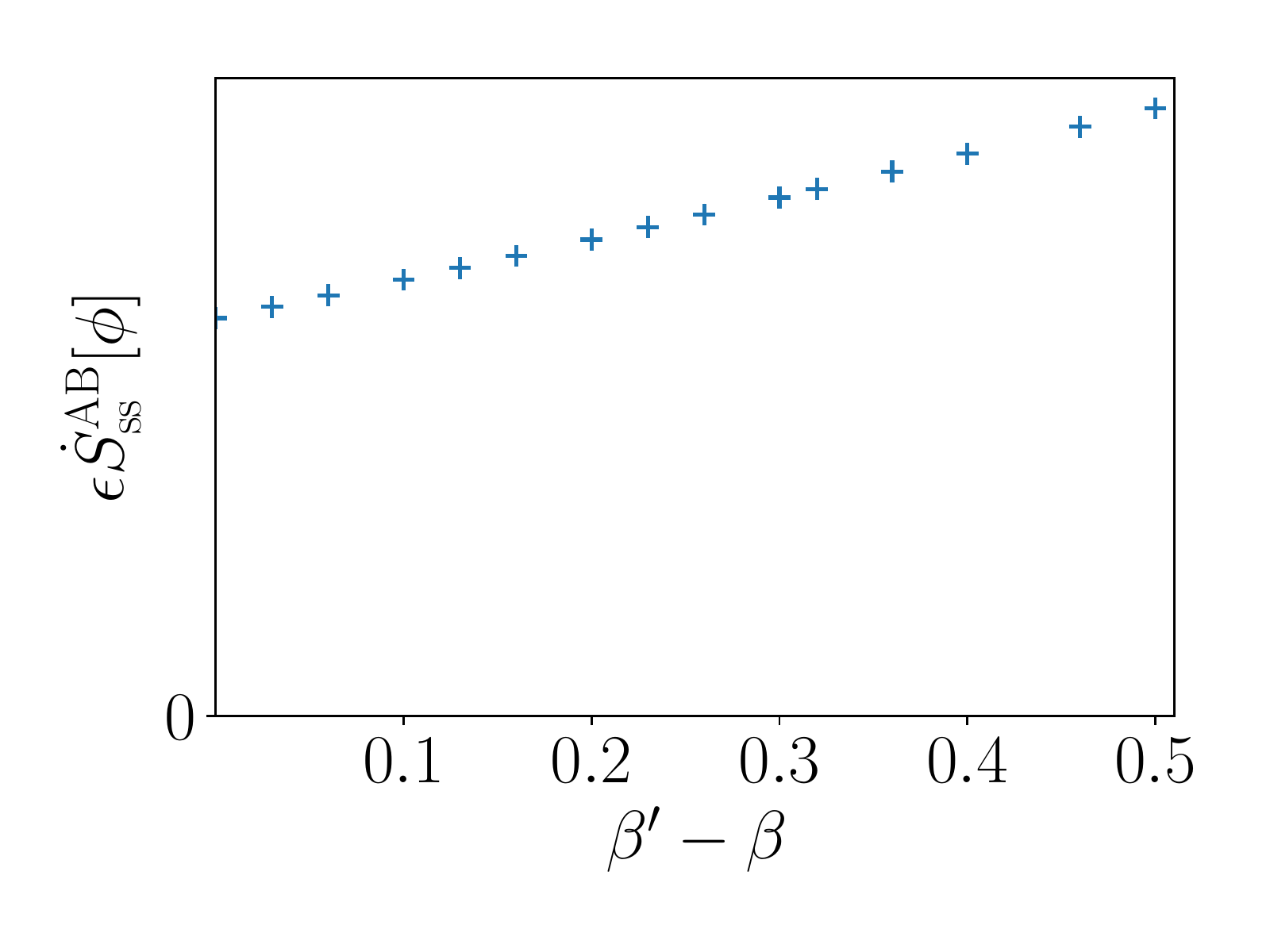}
\caption{}
\label{fig:epr_AB_tot}
\end{subfigure} 
\begin{subfigure}{0.55\textwidth}
\includegraphics[width=\textwidth]{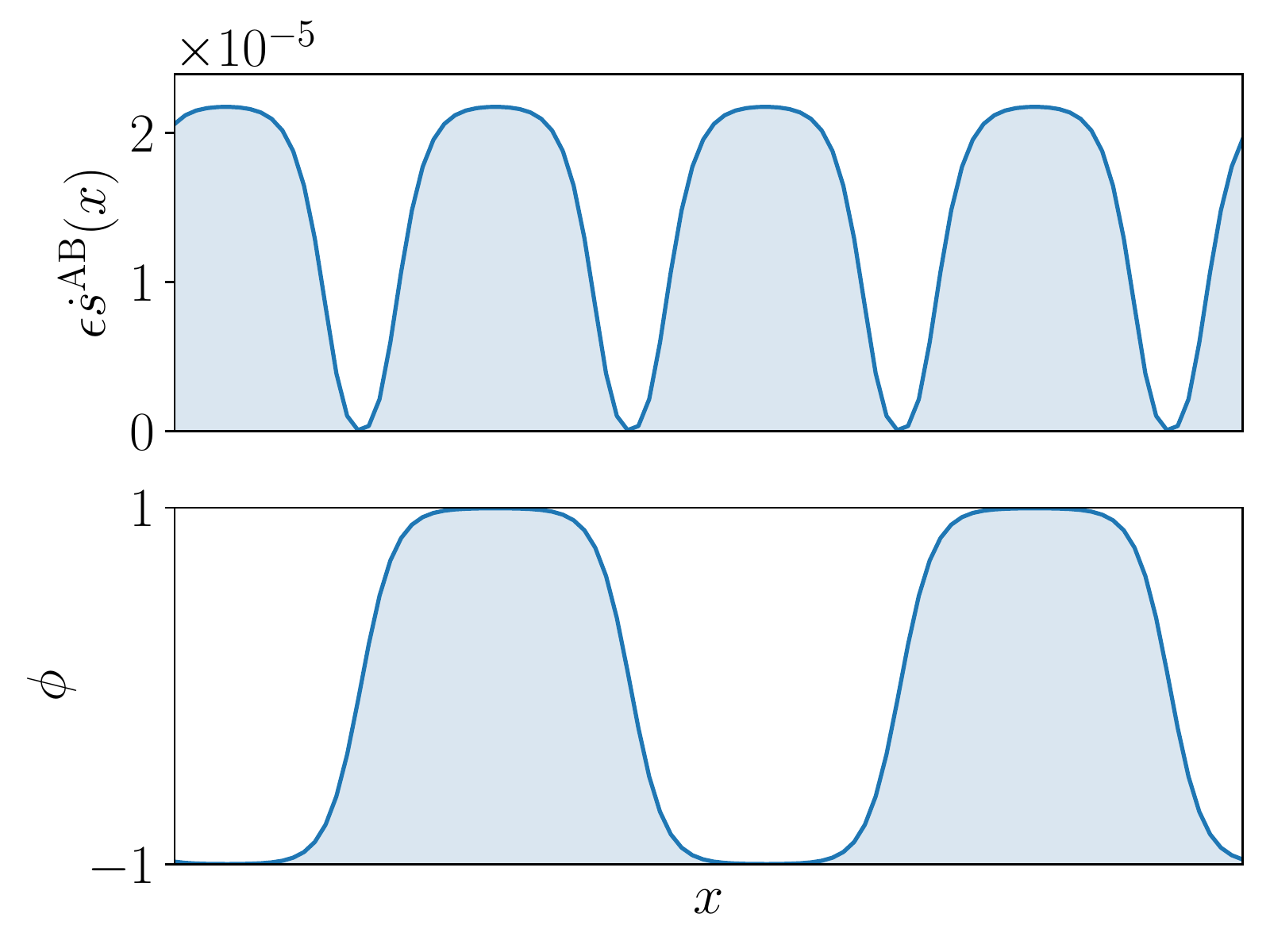}
\caption{}
\label{fig:epr_AB_local}
\end{subfigure} 
\caption{(A) Plot of $\epsilon \dot{S}_\mathrm{ss}^\mathrm{AB}[\phi]$ against $\beta' - \beta$. In contrast with Fig.\ \ref{fig:epr_phi}, the entropy production rate remains non-zero in the equilibrium subspace. (B) The upper panel shows the spatial decomposition of the entropy production rate $\dot{s}^\mathrm{AB}(x)$ and the lower panel shows the corresponding $\phi(x)$. Most of the entropy production is due to the Model A currents in the plateaus. The parameter values are $\alpha=\beta=\alpha'=1, \beta' = 1.1, \kappa=5, M_\mathrm{B} = 0.1, M_\mathrm{A} = 5 \times 10^{-6}$.}
\label{fig:epr_AB}
\end{figure}

The new entropy production rate $\dot{S}^\mathrm{AB}_\mathrm{ss}[\phi]$ is plotted in Fig.\ \ref{fig:epr_AB} for the same parameters as the numerical calculations that produced Fig.\ \ref{fig:epr_phi} in the previous section. We also retain the assumption that $\epsilon \ll 1$ and only compute expressions to lowest order in $\epsilon$, for easy comparison with the calculation of $\dot{S}_\mathrm{ss}$. Since both $\mu_\mathrm{A}, \bnabla \mu_\mathrm{B}$ are non-zero at the deterministic level, whereas $\mathcal{V}_\mathrm{ss}$ is of order $\epsilon$ (see equation (\ref{eq:quasipotential})), we can see that, to leading order in $\epsilon$, $Y_\mathrm{A, B}$ are only functions of the deterministic steady state $\phi^0_\mathrm{ss}$: $Y_\mathrm{A}^2 = M_\mathrm{A} ( \mu_\mathrm{A}(\phi^0_\mathrm{ss}) )^2 + O(\sqrt{\epsilon}) $ and $Y_\mathrm{B}^2 = M_\mathrm{B} (\bnabla \mu_\mathrm{B}(\phi^0_\mathrm{ss}))^2 + O(\sqrt{\epsilon}) $. As a result, the new entropy production rate $\dot{S}_\mathrm{ss}^\mathrm{AB}$, as well as its local decomposition $\dot{s}^\mathrm{AB}(\bx)$, are of order $1/\epsilon$, as opposed to $\dot{S}_\mathrm{ss}$ and $\dot{s}(\bx)$, which are both at most $O(\epsilon^0)$ as shown in equation (\ref{eq:local_phi_EPR}).  

The difference between the two EPRs is manifested more dramatically in the equilibrium subspace: $\mu_\mathrm{A}$ and $\bnabla \mu_\mathrm{B}$ both remain finite in this limit and so does $\dot{S}^\mathrm{AB}_\mathrm{ss}$, as shown in Fig.\ \ref{fig:epr_AB_tot}, whereas $\dot{S}_\mathrm{ss}$ vanishes as the special subspace is approached (see Fig.\  \ref{fig:epr_tot}). Near the equilibrium subspace, $\dot{S}_\mathrm{ss}^\mathrm{AB}$ appears to increase linearly with with the deviation $\beta' - \beta$, in contrast with the quadratic scaling observed for $\dot{S}_\mathrm{ss}$. This can be explained by expanding $\dot{S}_\mathrm{ss}^\mathrm{AB}$ for small $\Delta = \beta' - \beta$ as before. The deterministic steady state solution $\phi^0_\mathrm{ss}$ remains approximately the same across the range of $\Delta$ probed. Furthermore, as we only vary $\beta'$ but keep $\beta$ fixed, $(\bnabla \mu_\mathrm{B})$ stays constant while $\mu_\mathrm{A}(\Delta) = \mu_\mathrm{A}(\Delta = 0) + O(\Delta)$, leading to an $O(\Delta)$ piece in $Y_\mathrm{A}^2$ and hence $\dot{S}^\mathrm{AB}_\mathrm{ss} \propto \Delta $ upon spatial integration. 

Another interesting observation is that the local entropy production rate $\dot{s}^\mathrm{AB}(x)$ is more prominent at the plateaus, although the value remains finite at the interfaces, as shown in Fig.\ \ref{fig:epr_AB_local}. To explain this phenomenon, we start with examining the local Model A and Model B contributions separately:  $Y_\mathrm{A}^2 (= M_\mathrm{A} \mu_\mathrm{A}^2)$ is large at the plateaus, where the magnitude of the field $| \phi^0_\mathrm{ss} |$ is maximum, and goes to zero at the interfaces where $\phi^0_\mathrm{ss} = 0$; on the other hand, $Y_\mathrm{B}^2 (=  M_\mathrm{B} | \bnabla \mu_\mathrm{B} | ^2)$ is large at the interface, where the gradient is sharp, and small in the plateaus. Next, let the length scale of the pattern be $L$. In steady state, we have, approximately, $M_\mathrm{B} \mu_\mathrm{B}/ L^2 \sim M_\mathrm{A} \mu_\mathrm{A}$, from which we can deduce that $Y_\mathrm{A}^2/ Y_\mathrm{B}^2 \sim M_\mathrm{B}/ (M_\mathrm{A} L^2 )$. For the input parameters of our simulations, this ratio is much greater than 1. Thus the spatial decomposition $\dot{s}^\mathrm{AB}$ is dominated by Model A contributions, which accounts for the peaks at the plateaus. 

To conclude, in this section we presented an alternative way of tracking information for Model AB, and arrived at new expressions for the total steady state entropy production rate $\dot{S}^\mathrm{AB}_\mathrm{ss}$ and the local production rate $\dot{s}^\mathrm{AB}(\bx)$. Both quantities are markedly different from their counterparts in section \ref{entropy_modelab}, in terms of their behaviour in the equilibrium subspace, scaling with the deviation ($\beta' - \beta$) and the spatial profile of the local entropy production. Generically there is no pathway of conversion between the new entropy production rates and the ones in section \ref{entropy_production}, highlighting the importance of the information tracked. Recall that these calculations treat entropy production as an informatic quantity, and hence dependent on what variables are tracked; we are not attempting to calculate a physical heat production whose full elucidation would require tracking of {\em all} microscopic sources of dissipation underlying the model (see \cite{Markovich2020}).

\section{Conclusion} 

In this paper, we presented a method to compute the entropy production rate $\dot{S}$ for scalar Langevin systems with additive noise, as a quantitative measure of the extent to which the time reversal symmetry is broken at the macroscopic scale. We discussed in detail the EPR of equilibrium systems and their non-equilibrium extensions, as well as a case study of Model AB that describes systems with mismatched conservative and non-conservative dynamics. 

Following the work of Seifert \cite{seifert, seifert2005entropy} in stochastic thermodynamics, we defined the field-theoretic entropy production $\Delta S$ for a trajectory as the difference between the rate function of the forward path and that of the reversed path. The entropy production rate is subsequently obtained by differentiating $\Delta S$ with respect to time. It is often convenient to split $\Delta S$ into an internal piece $\Delta S_\mathrm{Int}$ and an external piece $\Delta S_\mathrm{Ext}$, before taking the time derivatives of both and summing the internal and external contributions to give the total rate of change.  The internal entropy production is a ``surface term'' in time, accounting for the difference in the Gibbs entropies of the initial and final state. The external part is the contribution from the action. Though $\Delta S_\mathrm{Ext}$ can be related to the heat production in simple particle-based systems, its physical interpretation in general systems is less clear. 

In field theories, the probability of a trajectory in configuration space, and hence its rate function and the entropy production rate, depends on the macroscopic variables used to label the path. We first explored the natural choice of tracking $(\partial_t \phi, \phi)$ trajectories. In steady state, we found that the resulting entropy production rate $\dot{S}_\mathrm{ss}[\phi]$ of every field configuration is non-negative and there also exists a spatial decomposition $\dot{s}(\bx)$ that is guaranteed to be locally non-negative. Both are only function of the time-antisymmetric component of the dynamics $F_\mathrm{a} = F + K \delta_\phi \mathcal{V}$, which is another measure of the deviation from equilibrium, analogous to the antisymmetric currents in Macroscopic Fluctuation Theory. In equilibrium, both $\dot{S}_\mathrm{ss}$ and $\dot{s}(\bx)$ vanish as expected from the principle of detailed balance. For Model-B type systems with mass conservation, the local entropy production is $(\epsilon \epsilon M)^{-1} | \widetilde{\boldsymbol{J}}_\mathrm{a}|^2$ where $\widetilde{\boldsymbol{J}}_\mathrm{a}$ is the pure gradient part of the antisymmetric current, as any curl-part does not show in the $\phi$-evolution and therefore cannot contribute to the entropy production. 

In practice, the computation of the antisymmetric dynamics $F_\mathrm{a}$ and therefore the steady state EPR requires the knowledge of the quasipotential $\mathcal{V}_\mathrm{ss}$, which is only known exactly for solvable systems. To make progress, we performed a perturbative expansion in the noise magnitude $\epsilon$ that only requires the (numerical or analytical) solution of the deterministic equation. In steady state, both the total EPR and its spatial decomposition are at most of order $\epsilon^0$. In the special case where the noise kernel $K$ and the Jacobian $A$ are simultaneously diagonalisable, both $\dot{S}_\mathrm{ss}$ and $\dot{s}$ are $O(\epsilon)$, echoing the findings of Nardini et al.\ \cite{nardini2017entropy}. For Model AB, we found that the total steady state EPR vanishes in the equilibrium subspace and scales quadratically with the deviation $\beta' - \beta$ near the subspace. The spatial decomposition $\dot{s}(x)$ is positive everywhere with small peaks at the interfaces. 

To understand the effect of tracking different macroscopic variables, we then computed the EPR of Model AB with additional information on contributions from the Model A and Model B sectors. The new entropy production rate $\dot{S}^\mathrm{AB}$ is of order $\epsilon^{-1}$, reflecting the fact that once the conservative and non-conservative parts are tracked separately, the dynamics is irreversible in time even at the deterministic level. In steady state, the EPR $\dot{S}^\mathrm{AB}_\mathrm{ss}$ of each field configuration is also non-negative and we can identify a non-negative spatial decomposition that constitutes of a Model A contribution related to the local reactions and a Model B contribution from the macroscopic currents. Both the new steady state EPR $\dot{S}^\mathrm{AB}_\mathrm{ss}$  and its spatial decomposition $\dot{s}^\mathrm{AB}$ show qualitatively different phenomena to their counterparts in the previous case: $\dot{S}^\mathrm{AB}_\mathrm{ss}$ remains finite in the special subspace and increases linearly with $\beta' - \beta$; $\dot{s}^\mathrm{AB}$ exhibits large peaks at the plateaus instead of the interfaces.

While in this work we mainly focused on the small noise expansion, there is a lot of potential for interesting studies on whether the spatial decomposition will change as the noise amplitude increases, as there is no reason to believe that the terms that are higher order in $\epsilon$ have the same spatial distribution as the leading order term. In addition, our formalism rests mainly on the additivity of noise within the Langevin framework and not the scalar character of the fields involved (although we restricted to this case for simplicity of notation). Accordingly it can be easily extended to vectorial or tensorial systems with applications to swarming active matter and active liquid crystals. This could allow a deeper understanding of entropy production in these wider classes of active systems, whose analysis using stochastic thermodynamics has recently been initiated  \cite{borthne, Markovich2020} 

\section*{Acknowledgements:}
We thank Yongjoo Baek, Rob Jack, Cesare Nardini and \'{E}tienne Fodor for valuable discussions. YIL thanks the Cambridge Trust and the Jardine Foundation for a PhD studentship. This work was funded in part by by the European Research Council under the Horizon 2020 Programme, ERC grant agreement number 740269. MEC is funded by the Royal Society.

\begin{appendix}

\section{Connections to Macroscopic Fluctuation Theory}
 \label{ap:mft}
Our decomposition of $F$ into the symmetric and antisymmetric parts $F_\mathrm{a, s}$ in equation (\ref{eq:f_a})  is similar to the decomposition of the diffusive current $\boldsymbol{J}$ in Macroscopic Fluctuation Theory (MFT) literature \cite{mft}: $\boldsymbol{J}_\mathrm{s} = - \bnabla \partial_\phi \mathcal{V}_\mathrm{ss}, \boldsymbol{J}_\mathrm{a} = \boldsymbol{J} + \bnabla \partial_\phi \mathcal{V}_\mathrm{ss}$. While our derivation in section (\ref{f_a}) relies on the reversal of the Fokker-Planck equation, in MFT the ``adjoint dynamics'' is defined for the action of each path: the probability of observing the reversed path in the MFT-``adjoint dynamics'' is the same as the probability of the forward path in the forward dynamics. In this appendix, we ask the following question: is the adjoint dynamics in our definition equivalent to the MFT definition? 

Denote the rate function of the adjoint dynamics as $\mathbb{R}^\mathrm{R}$, the rate function of the reversed path is, 
\begin{equation*}
\eqalign{
\fl \mathbb{R}^\mathrm{R} \left [ (\phi^\mathrm{R}, \partial_t \phi^\mathrm{R})_{t \in [0, T]} \right ] 
= \frac{1}{4} \int \mathrm{d} t \, \left [ (- \partial_t \phi - F^\mathrm{R} ) ^\dagger K^{-1} ( - \partial_t \phi - F^R ) + 2 \epsilon \mathrm{Tr} \left ( \delta_\phi F^\mathrm{R} \right ) \right ]  \\ 
\qquad \quad + \mathcal{V}[ \phi(T), T] 
}
\end{equation*}
where $F^\mathrm{R}$ ($= - F - 2 K \delta_\phi \mathcal{V} $) is the reversed dynamics we found in section (\ref{f_a}) by reversing the Fokker-Planck equation. If $\mathbb{R}^\mathrm{R} \left [ (\phi^\mathrm{R}, \partial_t \phi^\mathrm{R})_{t \in [0, T]} \right ]  = \mathbb{R} \left [ (\phi, \partial_t \phi)_{t \in [0, T]} \right ]$, our definition of the adjoint dynamics gives path-wise reversal and therefore would be equivalent to the MFT definition. Taking the difference of the forward and the backward rate functions, we obtain
\begin{equation*}
\eqalign{  
\fl \mathbb{R}  \left [ (\phi, \partial_t \phi)_{t \in [0, T]} \right ]  - \mathbb{R}^\mathrm{R} \left [ (\phi^\mathrm{R}, \partial_t \phi^\mathrm{R})_{t \in [0, T]} \right ]  =
- \frac{1}{2} \int \mathrm{d} t \,   \partial_t \phi^\dagger  K^{-1}  ( F + F^\mathrm{R} )  \\
 +   \mathcal{V}[ \phi(0), 0] - \mathcal{V}[ \phi(T), T]  \\
 + \frac{1}{4} \int \mathrm{d} t \,  \left [ F^\dagger K^{-1} F -  (F^\mathrm{R})^\dagger K^{-1} F^\mathrm{R} \right ]   \\
 + \frac{1}{2} \epsilon \int \mathrm{d} t \, \mathrm{Tr} \left [ \delta_\phi F - \delta_\phi F^\mathrm{R} \right ] 
} 
\end{equation*}
Dividing by $T$ and taking the limit $T \rightarrow 0$ on both sides, 
\begin{equation*}
\eqalign{  
\fl \partial_t  \mathbb{R} - \partial_t  \mathbb{R}^\mathrm{R}  = - \frac{1}{2} \partial_t \phi^\dagger K^{-1}  ( F + F^\mathrm{R} + 2 K \frac{\delta \mathcal{V} }{\delta \phi }) + \frac{1}{4}  \left ( F^\dagger K^{-1} F -  (F^\mathrm{R})^\dagger K^{-1} F^\mathrm{R} \right )\\
- \partial_t \mathcal{V} + \frac{1}{2} \epsilon \mathrm{Tr} \left [ \delta_\phi F - \delta_\phi F^\mathrm{R} \right ]
}  
\label{eq:pathwise_reversal}
\end{equation*}
Substituting in $F^\mathrm{R} = - F - 2K \delta_\phi \mathcal{V}$, 
\begin{equation*}
\fl \partial_t  \mathbb{R} - \partial_t  \mathbb{R}^\mathrm{R}  = \frac{1}{4} \left ( F^\dagger K^{-1} F - ( F^\mathrm{R})^\dagger K^{-1} F^\mathrm{R} \right ) -  \partial_t \mathcal{V}  + \frac{1}{2} \epsilon \mathrm{Tr} \left [ \delta_\phi F - \delta_\phi F^\mathrm{R} \right ]
\end{equation*}
Or in terms of the symmetric and the antisymmetric dynamics defined as $F_\mathrm{s, a} = \left ( F \pm F^\mathrm{R} \right )/2$, 
\begin{equation*} 
\partial_t  \mathbb{R} - \partial_t  \mathbb{R}^\mathrm{R}  = F_\mathrm{s}^\dagger K^{-1} F_\mathrm{a} - \partial_t \mathcal{V} + \epsilon \mathrm{Tr} \left[  \delta_\phi F_\mathrm{a} \right ] 
\end{equation*}  
In MFT, the adjoint dynamics is defined for the steady state only ($\partial_t \mathcal{V} =0$) and in the limit $\epsilon \rightarrow 0$. This gives $\partial_t  \mathbb{R} - \partial_t  \mathbb{R}^\mathrm{R}  = F_\mathrm{s}^\dagger K^{-1} F_\mathrm{a} $, so if $F_\mathrm{s}$ and $F_\mathrm{a}$ are orthogonal, the adjoint dynamics becomes the path-wise reversal of the forward dynamics and the two definitions become equivalent. Thus we have shown equivalence of the definition of the adjoint dynamics used in this paper to the previous definition used in MFT subject to the orthogonality of $F_\mathrm{s}$ and $F_\mathrm{a}$. It lies beyond our present scope to establish when such orthogonality is actually present beyond the diffusive systems in \cite{mft}, corresponding to Model B and its extensions addressed in section \ref{relax}. 

\section{Expectation values with Stratonovich discretisation} \label{ap:exp}
In this appendix, we will define and evaluate the conditional expectations $\expval{ \partial_t \phi | \phi, t }$ and $\expval{ \partial_t \phi_\mathrm{A, B} | \phi, t}$ that are essential in averaging over the time derivatives to obtain an expression for the EPR of a field configuration $\phi$ in section \ref{phi_dot_epr} and \ref{phi_dot_a_epr}. We will only demonstrate the calculation explicitly for a simple one dimensional case, though generalisation to field theories should be relatively straight forward. 

Consider a single-variable stochastic differential equation (SDE) with non-multiplicative noise, 
\begin{equation*}
\dot{x} = f(x) + \sqrt{2 \epsilon K} \Lambda
\end{equation*}
where $\Lambda$ is a unit white noise. This equation is only well-defined once the discretisation scheme is specified, but as the noise is non-multiplicative all discretisations are equivalent. Without loss of generality we choose the endpoint discretisation: for a set of discrete time steps $\{ t_i \}$ with spacing $\Delta t$, the discretised SDE is 
\begin{equation*}
\Delta x (t_i) = f[ x(t_{i+1}) ] \Delta t + \Delta W (t_i)
\end{equation*}
where $\Delta x(t_i) = x(t_{i+1}) - x(t_i)$ and $W$ is a Wiener process with correlation $\expval{ \Delta W(t_i) \Delta W(t_j)} = 2 \epsilon K \Delta t \delta_{ij}$. Note that the discretisation of the SDE is a separate choice from that of the path integral. For the purpose of entropy production, we always use Strato-discretised action functional but that does not limit our choice of discretisation for the SDE. 

The conditional expectation $\expval{\dot{x} | x, t}$ in section \ref{phi_dot_epr} is the expectation value of $\dot{x}$ at time $t$ given that a fixed value of x at time $t$. The $\dot{x}$ here can traced back to the $\dot{x}$ in the action functional $\mathbb{A}$, which follows the Stratonovich (midpoint) scheme. Thus the discretised version of the conditional expectation is, 
\begin{equation*}
\expval{\dot{x} | x, t} = \mathbb{E} \left [ \lim_{\Delta t \rightarrow 0} \frac{ \Delta x(t_{i-1}) + \Delta x(t_i)}{2\Delta t} \middle | x(t_i) = x \right ] 
\end{equation*}
The rest of this derivation is presented in Seifert's papers \cite{seifert2005entropy} and we will not repeat here. He showed that $\expval{\dot{x} | x, t} = f(x) - \epsilon K \partial_x \log P(x, t) = f(x) + K \partial_x V(x, t)$, where $V(x, t)$ is the quasipotential defined in the same way as for fields: $P(x, t) \propto \exp( - \epsilon^{-1} V(x, t))$. 

Next, we consider the one dimensional version of Model AB, where there are two separate contributions to the time derivative with independent noises. As before, we choose endpoint discretisation though all choices are equivalent, 
\begin{equation*}
\eqalign{ 
\Delta x (t_i) &= \Delta x_\mathrm{A} (t_i)+ \Delta x_\mathrm{B} (t_i) \\
\Delta x_\mathrm{A} ( t_i ) & = f_\mathrm{A}[ x(t_{i+1}) ] \Delta t +  \Delta W_\mathrm{A}(t_i) \\
\Delta x_\mathrm{B} ( t_i ) & = f_\mathrm{B}[ x(t_{i+1}) ] \Delta t +  \Delta W_\mathrm{B}(t_i) 
} 
\end{equation*}
where $W_\mathrm{A, B}$ are independent Wiener processes with $\expval{\Delta W_\mathrm{A, B} (t_i) \Delta W_\mathrm{A, B} (t_j) } = 2 \epsilon K_\mathrm{A, B} \Delta t \delta_{ij}$. The one dimensional version of the conditional expectation in section \ref{phi_dot_a_epr} is $\expval{ \dot{x}_\mathrm{A} | x, t}$, formally defined as
\begin{equation*}
\expval{\dot{x}_\mathrm{A} | x, t} = \mathbb{E} \left [ \lim_{\Delta t \rightarrow 0} \frac{ \Delta x_\mathrm{A}(t_{i-1}) + \Delta x_\mathrm{A}(t_i)}{2\Delta t} \middle | x(t_i) = x \right ] 
\end{equation*}
Our calculation for the conditional expectation mainly follows the line of Seifert's calculation \cite{seifert2005entropy}. First, by linearity of the conditional expectation, we can evaluate the following before taking the limit of $\Delta t \rightarrow 0$, 
\begin{equation*}
\eqalign{  
\fl \mathbb{E} \left [  \Delta x_\mathrm{A}(t_{i-1}) + \Delta x_\mathrm{A}(t_i) | x(t_i) = x \right ]  &=  \mathbb{E} \left [ \Delta x_\mathrm{A}(t_{i-1}) | x(t_i) \right ] + \mathbb{E} \left [ \Delta x_\mathrm{A}(t_i) | x(t_i) = x \right ] \\
&=   \mathbb{E} \left [ f_\mathrm{A}[x_{i-1}] \Delta t + \Delta W_\mathrm{A}(t_{i-1}) | x(t_i) = x \right ] \\
&\quad + \mathbb{E} \left [ f_\mathrm{A}[x(t_i)] \Delta t + \Delta W_\mathrm{A}(t_i) | x(t_i) = x \right ] 
} 
\end{equation*}
The second term is a forward time conditional expectation. Using the non-anticipating property of the Wiener process, 
\begin{equation*}
\eqalign{  
\fl \mathbb{E} \left [ f_\mathrm{A}[ x(t_{i+1})] \Delta t + \Delta W_\mathrm{A}(t_i) | x(t_i) = x \right ] & = \mathbb{E} \left [ f_\mathrm{A}[x(t_{i+1})] \Delta t | x(t_i) = x \right ] \\
& = \mathbb{E} \left [ f_\mathrm{A}[x(t_i)] \Delta t  + O\left (\Delta t^{3/2}\right ) |x(t_i) = x \right ] \\
 & = f_\mathrm{A}(x) \Delta t +  O\left (\Delta t^{3/2}\right ) 
} 
\end{equation*}
The first term is trickier as $x(t_i)$ is not independent of $\Delta W_\mathrm{A}(t_{i-1})$. Writing the conditional expectation of the noise term as an explicit integral over the probabilities, 
\begin{equation*}
\fl \mathbb{E} \left [ f_\mathrm{A}[x(t_i)] \Delta t + \Delta W_\mathrm{A}(t_{i-1}) | x(t_i) = x \right ]  
=  f_\mathrm{A} (x)+ \int \mathrm{d} \xi  \,  \xi P \left (\Delta W_\mathrm{A}(t_{i-1}) = \xi \middle | x(t_i) = x \right ) 
\end{equation*}
Omitting the time label on $\Delta W_\mathrm{A}$ for brevity, and using Bayes' theorem for the conditional probability, 
\begin{equation*}
\eqalign{  
 \fl \mathbb{E} \left [  \Delta W_\mathrm{A}| x(t_i) = x \right ]  &= \int \mathrm{d} \xi P \left (x(t_i) = x \middle | \Delta W_\mathrm{A} = \xi \right ) \frac{P(\Delta W_\mathrm{A} = \xi) }{P(x(t_i) = x)} \\
&= \int \mathrm{d}\xi \mathrm{d} \eta \, P(x(t_i) = x | \Delta W_\mathrm{A} = \xi, x(t_{i-1}) = \eta) \frac{P(\Delta W_\mathrm{A} = \xi) P(x(t_{i-1}) = \eta ) }{P(x(t_i) = x) } \\
} 
\end{equation*}
Inspection of the conditional probability $P(x(t_i) = x | \Delta W_\mathrm{A} = \xi, x(t_{i-1}) = \eta)$ reveals that it is equivalent to the probability of the difference 
\begin{equation*}
\fl P(x(t_i) = x | \Delta W_\mathrm{A} = \xi, x(t_{i-1}) = \eta ) 
= P( \Delta W_\mathrm{B} = x - \eta - \xi - f_\mathrm{A}(x) \Delta t - f_\mathrm{B}(x) \Delta t ) 
\equiv P(\Delta W_\mathrm{B} = \zeta ) 
\end{equation*}
where we defined a new variable $\zeta$ at the last equality. Next, we simplify the following fraction in the integrand by expanding in small $\Delta x$, 
\begin{equation*}
\frac{P(x(t_{i-1}) = \eta ) }{P(x(t_i) = x) } = 1 - (x - \eta )\partial_x \log P(x, t_i)
\end{equation*}
Changing the integration variable from $\eta$ to $\zeta$ and combining the terms, 
\begin{equation*}
\eqalign{  
\fl \mathbb{E} \left [  \Delta W_\mathrm{A} | x(t_i) = x \right ] &=  \int \mathrm{d} \xi \mathrm{d} \zeta \, \xi  P(\Delta W_\mathrm{B} = \zeta) P(\Delta W_\mathrm{A} = \xi ) \left [  1- ( \zeta + \xi + f_\mathrm{A} (x) \Delta t + f_\mathrm{B} (x) \Delta t ) \partial_x \log P \right ] \\
&= -  \int \mathrm{d} \xi \, \xi^2 P(\Delta W_\mathrm{A} = \xi) \partial_x \log P + O\left (\Delta t^{3/2}\right ) \\
& = - 2 \epsilon K_\mathrm{A} \Delta t \partial_x \log P + O\left (\Delta t^{3/2}\right ) \\
} 
\end{equation*}
where we have used the correlation for the Wiener process $W_\mathrm{A}$ in the last line. 
Adding the terms together, the conditional expectation is
\begin{equation*}
\mathbb{E} \left [  \Delta x_\mathrm{A}(t_{i-1}) + \Delta x_\mathrm{A}(t_i) | x(t_i) = x \right ]   = 2 f(x) \Delta t - 2 \epsilon K_\mathrm{A} \partial_x \log P(x, t_i) 
\end{equation*}
Dividing both sides by $\Delta t$ and taking the limit $\Delta t \rightarrow 0$, we obtain the desired conditional expectation, 
\begin{equation*} 
\mathbb{E} \left [ \dot{x}_\mathrm{A}(t) | x(t) = x \right ] = f(x) - \epsilon K_\mathrm{A} \partial_x \log P(x, t)  = f(x) + K_\mathrm{A} \partial_x V(x, t)
\end{equation*} 
This concludes the calculation the one dimensional version of the conditional expectation values mentioned in section \ref{phi_dot_a_epr}, and we refer back to the main text for how it is used to calculate the entropy production rate of $(\phi, \partial_t \phi_\mathrm{A}, \partial_t \phi_\mathrm{B})$ trajectories. 

\section{Entropy production with continuous symmetry} 
\label{ap:symmetry}
This appendix discusses the details of the calculation of the local entropy production in section \ref{small_noise} in the case of a spontaneously broken continuous symmetry in the system. 

In the presence of the broken symmetry, there exists a Goldstone mode $v$: an eigenvector of the matrix $A$ with zero eigenvalue. In our case, with periodic boundary conditions, our system possesses translational symmetry (invariance under $\phi(x) \rightarrow \phi(x+\delta x)$). In fact, the Goldstone mode $v$ equals $\partial_x \phi^0_\mathrm{ss}$, where $\phi^0_\mathrm{ss}$ is the deterministic steady state solution. Another consequence is that the quasipotential $\mathcal{V}_\mathrm{ss}$ also has the same symmetry. Recall from equation (\ref{eq:quasipotential}) that $\mathcal{V}_\mathrm{ss}[\phi_1] = - \frac{\epsilon}{2} \phi_1^\dagger G \phi_1$ (without the degeneracy, $G$ is the inverse of the correlation function). $G$ also has $v$ as an eigenvector with eigenvalue zero. The projection operator $P = 1 - v v^\dagger$ defines the projection into the subspace without the Goldstone mode. We can immediately see that $AP = A$ and $G P = P$. 

Recall from equation (\ref{eq:local_phi_EPR}) that to lowest order in the noise parameter $\epsilon$, 
\begin{equation*}
\expval { \dot{s} } = \mathrm{Diag} \left [ E C E^\dagger \right ] + O(\epsilon^{1/2}) 
\end{equation*}
where $E = \sigma^{-1} A + \sigma^\dagger G$. Note that we also have $E P = E$: the Goldstone mode $v$ is also an eigenvector of $E$ with zero eigenvalue. Hence 
\begin{equation*}
\expval{\dot{S}} = \mathrm{Diag} \left [ E (P C P) E^\dagger \right ] + O( \epsilon^{1/2}) 
\end{equation*}
This implies that only correlations in the subspace contribute towards the total entropy production. Hence we only need to solve the reduced Lyapunov equation obtained by multiplying the full Lyapunov equation by $P$ on the left and right, 
\begin{equation}
(P A P) (P C P) + (P C P)(P A^\dagger P) = - 2 (P K P) 
\label{eq:subspace} 
\end{equation}

In practice, we add a small eigenvalue $\lambda$ in the Goldstone mode, and define $A_\lambda = A + \lambda v v^\dagger$. This makes the original Lyapunov equation well-defined and solvable. Let the corresponding solution of the $\lambda$- regularised equation be $C_\lambda$ such that 
\begin{equation*}
A_\lambda C_\lambda + C_\lambda A_\lambda^\dagger = - 2 K 
\end{equation*}
Multiplying the equation on the left and right by the projection $P$ gives
\begin{equation*}
P A_\lambda C_\lambda P + P C_\lambda A_\lambda^\dagger P = - 2 P K P 
\end{equation*}
Note that $P A_\lambda = P (A + \lambda v v^\dagger) = P A$ and $A P = A$, the above equation is equivalent to 
\begin{equation*}
(P A P) (P C_\lambda P) + (P C_\lambda P) ( P A^\dagger P) = - 2 P K P 
\end{equation*}
Comparing with equation (\ref{eq:subspace}) reveals that $P C_\lambda P = P C P$. 

Next we proceed to calculate $G$. The quadratic form of the quasipotential leads to the inverse relation in the subspace $P C P G = G P C P = P$. Use the same trick as before and let $\tilde{C}_\lambda = P C P + \lambda v v^\dagger$ (not the same as $C_\lambda$), we have 
\begin{equation*}
P = P \tilde{C}_\lambda^{-1} \tilde{C}_\lambda P = P \tilde{C}_\lambda^{-1} P P C P 
\end{equation*}
Thus we can identify $G = P \tilde{C}_\lambda^{-1} P$ and use it to calculate $E$. Note that $\lambda$ remains finite throughout the calculation. 

\section{Numerical implementations} 
\label{ap:num} 
In this section, we outline the numerical implementations of the small noise expansion of the entropy production rate for Model AB. In section \ref{entropy_modelab}, we performed the computations in Fourier space and only transform back to the real space at the last step to obtain the spatial decomposition. The Fourier transform applies to infinite continuous spatial domain but needs to be modified for numerical studies as the simulation box size is only finite and so is the number of grid points. This is because the discrete Jacobian matrix $A$ used in the weak noise expansion must the same as the Jacobian of the numerical integration of the deterministic PDE such that $A$ is negative definite (modulo Goldstone modes) as assumed. 

For numerical integration of PDEs, there are two methods of spatial discretisation for domains with periodic boundary conditions: finite difference method and pseudospectral method. Both methods converge for Model AB simulations in one dimension for a relatively small spatial domain, such is the case of interest here. In this paper, we use the former for simplicity though our analysis can be easily extended to the latter. In finite difference scheme, the Laplacian operator is represented by the following matrix, 
\begin{equation}
(\nabla^2)_{mn} =  \frac{1}{a^2} \left (- 2 \delta_{m,n} + \delta_{m, n-1} + \delta_{m, n+1} \right )
\end{equation}
where the indices wrap around and $a$ is the lattice spacing. In our simulations, we always rescale the parameters such that $a = 1$ so we will drop it from now on. The finite-difference Laplacian is diagonalised by the discrete Fourier Transform matrix $U_{mn} = L^{-1/2} \exp( - 2 \pi i mn/L)$ with diagonal elements $d_n = - 2 \left [  1 - \cos( 2 \pi n / L) \right ] $, where $L$ is the length of the domain (remember $a=1$). As a sanity check, we can see that for small $n$ and large $L$, $d_n \approx (2 \pi n/ L)^2 = q_n^2$, where $q_n$ are the discrete Fourier modes. This means that the eigenspace for the discrete Laplacian operator is the (discrete) Fourier space and we can use Fast Fourier Transform algorithms to go between the eigenspace and the real space. In addition, we can obtain the discrete Jacobian matrix $A$ by substituting $d_n$ for $q^2$ in equation (\ref{eq:jacobian}) and the noise kernel $K$ is diagonal in Fourier space with elements $K_n = M_\mathrm{A} + M_\mathrm{B} d_n$. The rest of the computation follows rather straight forwardly by using the results of \ref{ap:symmetry} to take care of the Goldstone mode and the Lyapunov equation can be solved using the Bartels-Stewart algorithm  \cite{Lyapunov}, which is part of many popular numerical libraries such as Scipy and LAPACK.

\end{appendix}

\section*{Bibliography} 

\bibliographystyle{unsrt}
\bibliography{Entropy}

\end{document}